\documentclass[aps,nofootinbib,showpacs,showkeys,preprint] {revtex4-1}

\usepackage{epsf,epsfig,subfigure,graphicx,amsmath,amssymb}
\usepackage{color}
\usepackage{float}
\usepackage{multirow}

\newcommand{\dis}[1]{\begin{equation}\begin{split}#1\end{split}\end{equation}}

\newcommand{\gev}{\,\textrm{GeV}}

\newcommand{\ie}{{\it i.e.}\ }

\usepackage{xcolor}

\begin{document}

\title{\large{\bf D0 dimuon charge asymmetry from $B_s$ system with $Z'$ couplings \\
and the recent LHCb result}}

\author{
Hyung Do Kim$^1$\footnote{hdkim@phya.snu.ac.kr}
Sung-Gi Kim$^2$\footnote{kimsg@indiana.edu} 
and 
Seodong Shin$^{1,2}$\footnote{sshin@phya.snu.ac.kr}
}
\affiliation{
$^1$CTP and Department of Physics and Astronomy, Seoul National University, Seoul 151-747, Korea\\
$^2$Physics Department, Indiana University, Bloomington, IN 47405, USA \\
}
\begin{abstract}
The D0 collaboration has announced the observation of the like-sign dimuon charge asymmetry since 2010, which has more than $3\sigma$ deviation from the Standard Model prediction. One of the promising explanation is considering the existence of flavor changing $Z'$ couplings to the $b$ and $s$ quarks which can contribute to the off-diagonal decay width in the $B_s - \bar{B}_s$ mixing. Model construction is highly constrained by the recent LHCb data of 1fb$^{-1}$ integrated luminosity . In this paper, we analyze the experimental constraints in constructing new physics models to explain the dimuon charge asymmetry from the CP violation of the $B_s$ system. We present limits on $Z'$ couplings and show that it is impossible to obtain the $1\sigma$ range of the dimuon charge asymmetry without the new contribution in the $B_d$ system. Even with arbitrary contribution in the $B_d$ system, the new couplings must be in the fine tuned region.

\end{abstract}

\maketitle


\section{Introduction}
\label{sec:intro}

The like-sign dimuon charge asymmetry from the semi-leptonic ($s\ell$) decay of $B_{s,d}$ meson is given by,
\dis{
A_{s\ell}^b=\frac{N^{++}-N^{--}}{N^{++}+N^{--}},
}
where $N^{++}$ corresponds to each $B$ hadron decaying semi-leptonically to $\mu^+ X$, and similarly $N^{--}$ to $\mu^- X$. In 2010, the D0 collaboration at the Tevatron announced the first observation of the large dimuon charge asymmetry, which deviated about 3.2$\sigma$ from what is expected in the Standard Model (SM) \cite{Abazov:2010hv}.  In 2011, the result from the analysis with 9 fb$^{-1}$ data was announced as \cite{Abazov:2011yk} 
\dis{
A_{s\ell}^b = - (7.87 \pm 1.72 \pm 0.93 ) \times 10^{-3} ,\label{eq:D0exp}
}
which has about $3.9 \sigma$ deviation from the SM prediction \cite{Abazov:2011yk},
\dis{
A_{s\ell}^{b \rm SM} = (-2.8^{+0.5}_{-0.6}) \times 10^{-4}.
}
To explain the observed asymmetry, we need additional sources of CP violation from the new physics (NP) beyond the SM in the $B_{s,d}$ mixing and/or decay. 

The contribution from each neutral $B^0$ and $B_s^0$ meson is parametrized by the flavor specific asymmetry 
\dis{
&a_{s\ell}^d \equiv \frac{\Gamma(\overline{B}_d \to \mu^+ X) - \Gamma(B_d \to \mu^- X)}{\Gamma(\overline{B}_d \to \mu^+ X) + \Gamma(B_d \to \mu^- X)}~, \\
&a_{s\ell}^s \equiv \frac{\Gamma(\overline{B}_s \to \mu^+ X) - \Gamma(B_s \to \mu^- X)}{\Gamma(\overline{B}_s \to \mu^+ X) + \Gamma(B_s \to \mu^- X)}~.
}
The fraction of each flavor specific asymmetry in the total asymmetry $A_{s\ell}^b$ at the Tevatron energy 1.96 TeV depends on the mean mixing probabilities and the production fractions of $B^0$ and $B^0_s$ mesons such that \cite{Abazov:2011yk}
\dis{
A_{s\ell}^b = (0.594\pm 0.022) a_{s\ell}^d + (0.406\pm 0.022) a_{s\ell}^s ~,
\label{eq:Aarelation}
}
which leads to $6:4$ production of the like-sign dimuons from the  $b\bar d(d\bar b)$ and $b\bar s(s\bar b)$ mesons.{\footnote{
This is different from the 2010 prediction of about $5:5$ production.}} 

Imposing the lower limits of the muon impact parameter (IP), it is possible to reduce the background dramatically, which is mainly from the long-lived charged mother particles of the muon and the anti-muon. In the 2011 data, the separation of the sample by the muon impact parameter provides the separate determination of $a_{s\ell}^d$ and $a_{s\ell}^s$ such that
\begin{eqnarray}
a_{s\ell}^s &=& - (18.1 \pm 10.6) \times 10^{-3}~, \label{eq:asls} \\
a_{s\ell}^d &=& - (1.2 \pm 5.2) \times 10^{-3}~,  \label{eq:asld}
\end{eqnarray}
where the SM predictions  using the SM fit of $|V_{ub}| = (3.56^{+0.15}_{-0.20}) \times 10^{-3}$ \cite{Lenz:2011ti} are
\begin{eqnarray}
a_{s\ell}^{s \rm SM} & = & (1.9 \pm 0.3) \times 10^{-5}, \\
a_{s\ell}^{d \rm SM} & = & -(4.1 \pm 0.6) \times 10^{-4}.
\end{eqnarray}
The separately determined $a_{s\ell}^s$ has about $1.7 \sigma$ deviation from the SM prediction
if $a_{s\ell}^d$ can be freely chosen to fit the data.  Similarly the $a_{s\ell}^d$ is within $1\sigma$ if $a_{s\ell}^s$ can be arbitrary. 
It should be noted however that in order $A^b_{s\ell}$ to be within 1$\sigma$ from its measured value, a large contribution from new physics in $a_{s\ell}^s$ is necessary as we see bellow.

\begin{figure}
\subfigure[\ Combined fit]{
\includegraphics[width=7.5cm]{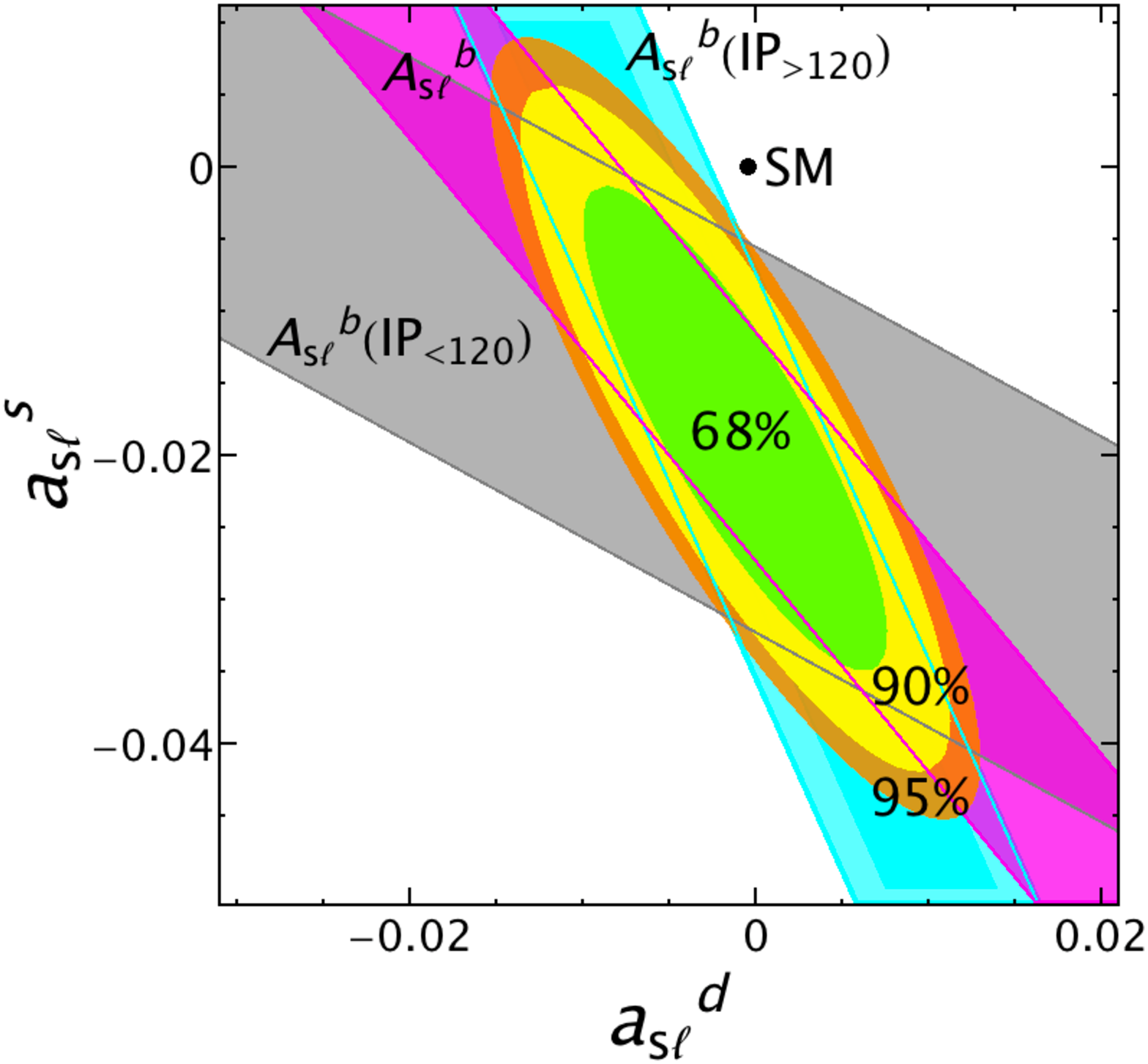}
}
\quad
\subfigure[\ Possible enhancement]{
\includegraphics[width=7.5cm]{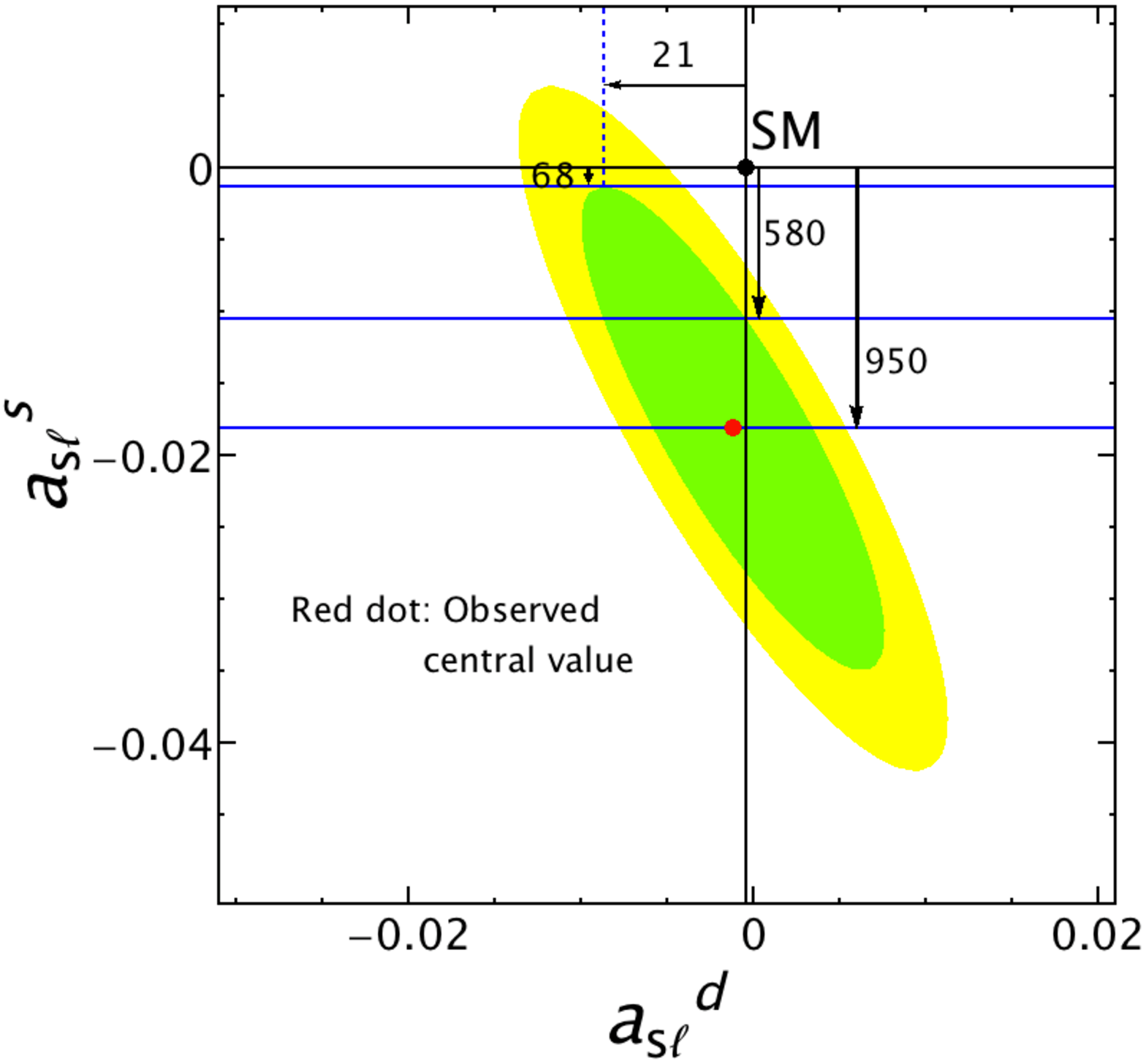}
}
\caption{We reproduced the measurements with different muon impact parameter (IP) selections according to \cite{Abazov:2011yk} in (a). The bands are the 90\% uncertainties on each individual measurement of IP$_{<120}$ (Gray), IP$_{>120}$ (Cyan), and the result without the IP cut (Purple) in Eq.(\ref{eq:D0exp}). The green (68\%), yellow (90\%), and orange (95\%) ellipses are obtained from the $\chi^2$-fit combining the measurements of IP$_{<120}$ and IP$_{>120}$ using the independent data sample. 
In Fig.\ref{fig:d0} (b), the red dot denotes the observed central values of $(a_{s\ell}^d, a_{s\ell}^s)$ while a black dot represents the SM predictions. 
We chose three representative values for $a^s_{s\ell}$ and one for $a^d_{s\ell}$, whose distances form the SM values are given in terms necessary enhancement of $-a_{s\ell}^s/(a_{s\ell}^s)^{\rm SM}$ and $a_{s\ell}^d/(a_{s\ell}^d)^{\rm SM}$.
}
\label{fig:d0}
\end{figure}

We reproduced the $\chi^2$-fit combining the impact parameter cut ($120\mu m$) analysis of IP$_{<120}$ and IP$_{>120}$ in Fig. \ref{fig:d0}. As seen in the figure, we need to consider both of $a_{s\ell}^d$ and $a_{s\ell}^s$ together in the 2D plane.
In this figure, we used the central values in the fraction of contribution by $a_{s\ell}^d$ and $a_{s\ell}^s$ in $A_{s\ell}^b$, shown in \cite{Abazov:2011yk}. 
In Fig. \ref{fig:d0} (b), the red dot denotes the observed central values of $(a_{s\ell}^d, a_{s\ell}^s)$ while the black dot represents the SM predictions. 
We chose three representative values for $a^s_{s\ell}$ and one for $a^d_{s\ell}$, whose distances form the SM values are given in terms necessary enhancement of $-a_{s\ell}^s/(a_{s\ell}^s)^{\rm SM}$ and $a_{s\ell}^d/(a_{s\ell}^d)^{\rm SM}$. 
First of all, if we allow arbitrarily new physics in $a_{s\ell}^d$, we still need at least 68 times bigger size of $a_{s\ell}^s$ to explain the asymmetry within $1\sigma$ where $a_{s\ell}^d/(a_{s\ell}^d)^{\rm SM} = 21$ at the $1\sigma$ boundary. 
Without NP contribution to $a_{s\ell}^d$, the enhancement $-a_{s\ell}^s/(a_{s\ell}^s)^{\rm SM}> 580$ is needed to explain the asymmetry within $1\sigma$. 
To be at the central point in which the $\chi^2$ fit is the best, $a_{s\ell}^s$ should be enhanced by a factor of 950 while small enhancement in $a_{s\ell}^d/(a_{s\ell}^d)^{\rm SM} < 3$ is enough. 
Therefore, we conclude that the observed value of $A^b_{s\ell}$ requires large NP contribution in $a_{s\ell}^s$.{\footnote{If we aim for the asymmetry within 90\% confidence region ($1.65\sigma$) or $2\sigma$ region, the observation result can be achieved only by the enhancement in $a_{s\ell}^d$ without having any contribution in $a_{s\ell}^s$.}} 
We take three points as references.
\dis{
(a_{s\ell}^d/(a_{s\ell}^d)^{\rm SM}, \,a_{s\ell}^s/(a_{s\ell}^s)^{\rm SM}) = (21,-68),~(1,-580),~(1,-950)~,
\label{eq:base}
}
where the small enhancement in $a_{s\ell}^d$ at the third reference point is not considered.

As a promising example explaining the large dimuon charge asymmetry, the $Z'$ scenario with both flavor diagonal and off-diagonal couplings has been analyzed \cite{Kim:2010gx,Alok:2010ij}. In this paper, we study the validity of $Z'$ boson explanation satisfying Eq.\eqref{eq:base} by checking recently updated experimental constraints from the $B$/$B_s$ meson decays and mixing \footnote{The electroweak precision test results can also provide strong constraints when the mixing of the $Z'$ and $Z$ boson exists \cite{Langacker:2000ju,Kim:2011xv}, while we do not consider such effect here.}. Especially, the recent LHCb results provide very strong bounds. We study the operators $(\bar{s}_X \gamma^\mu b_X)(\bar{\tau}_Y \gamma_\mu \tau_Y)$ or $(\bar{s}_X \gamma^\mu b_X)(\bar{c}_Y \gamma_\mu c_Y)$ (where $X, Y = L, R$), because NP contribution to $b \to s \tau^+ \tau^-$ is weakly constrained  from the Br($B_s \to \tau^+ \tau^-$) \cite{Bauer:2010dga}, and the effect on $a^s_{s\ell}$ from $(\bar{s}_X \gamma^\mu b_X)(\bar{c}_Y \gamma_\mu c_Y)$  can be enhanced by the interference with the $W$ boson exchange \cite{Alok:2010ij}. Unlike other papers, we present our results in terms of the actual $Z'$ couplings for a fixed $Z'$ mass, $M_{Z'}$. (Readers can simply rescale constraints on the couplings for different value of $M_{Z'}$.) Therefore, it will be easy to see the feasibility of realizing allowed space of $Z'$ couplings from the view point of model building that we don't discuss in this paper. The effective set-up only considering the $Z'\bar{s}b$ and $Z' \tau^+ \tau^-$ ($Z' c \bar{c}$) couplings is used, regardless of their theoretical origins. 

While we were in the completion of our work, a similar analysis for the operator of $(\bar{s}_X \gamma^\mu b_X)(\bar{c}_Y \gamma_\mu c_Y)$ was appeared \cite{Li:2012xc}. They chose special cases either one of the couplings of  $Z' c_L \bar{c}_L$ and $Z' c_R \bar{c}_R$ is turned off or they are set to equal. Comparing to this simplification, our analysis deals with general case with more systematic approach. By doing this, we point out the $Z' c \bar{c}$ couplings must be (almost) axial vector-like from the constraint $B_s \to J/\psi~\phi$ and quantitatively see how much the axial vector relation can be violated by combining other experimental bounds. For the $Z' \tau^+ \tau^-$ couplings, we note that our analysis includes the constraint from $b \to s \nu \bar{\nu}$ that has not been discussed in the preceding studies.

This paper is organized as follows. We provide a summary review of the $Z'$ explanation on the dimuon charge asymmetry in Sec. \ref{sec:dimuon}. Then, we analyze the current experimental bounds constraining the NP model construction explaining the asymmetry and apply the bounds to the $Z'$ properties in Sec. \ref{sec:ex}. The experimental results we will analyze contain the measurements of the mass difference $\Delta M_s$ and the width difference $\Delta \Gamma_s$ after the mixing. We also included the bounds from the CP violating phase $\phi_s^{J/\psi\,\phi}$ of the $B_s \to J/\psi\,\phi$ process, the inclusive $b \to s \nu \bar{\nu}$, and the $\sin2\beta$ from the golden plate $B \to J/\psi K_S$. In Sec. \ref{sec:gtata} and \ref{sec:gcc}, we directly obtain the combined constraint on the $Z'$ model parameters in the models with the $Z'\tau^+ \tau^-$ coupling and the $Z' c\bar{c}$ coupling, respectively. Finally, we give the conclusions in Sec. \ref{sec:conclusions}.

\section{The like-sign dimuon charge asymmetry}
\label{sec:dimuon}

The $B_q - \overline{B}_q$ oscillations for $q=s,d$ are described by a Schr\"odinger equation
\begin{eqnarray}
i \frac{\text{d}}{\text{d} t} \left( \begin{array}{c} \vert B^0 \rangle \\
  \vert \overline{B}^{\,0}  \rangle
  \end{array}  \right) = \left( M - i \frac{\Gamma}{2}\right) \left( \begin{array}{c} \vert B^0 \rangle \\
  \vert \overline{B}^{\,0} \rangle
  \end{array}  \right) ,
\end{eqnarray}
where $M$ and $\Gamma$ are the $2 \times 2$ Hermitian  mass and decay matrices, which are dispersive and absorptive parts in the time dependent mixing respectively. The differences of masses and widths of the physical eigenstates are given by the off-diagonal elements as  \cite{Lenz:2006hd}
\begin{eqnarray}
\Delta M_q = 2 |M_{12}^q| ~, \hspace{0.5cm} \Delta \Gamma_q = 2 |\Gamma_{12}^q| \cos\phi_q~,
\label{eq:M_12_and_DeltaM}
\end{eqnarray}
up to numerically irrelevant corrections of order $m_b^2 / M_W^2$ as long as $\Delta M \gg \Delta \Gamma$ for $B_q$ meson system. The CP phase difference between these quantities is defined as
\begin{eqnarray}
\phi_q = \mbox{Arg.}\left(-\frac{M_{12}^q}{\Gamma_{12}^q}\right) ,
\end{eqnarray}
where the SM contribution to this angle is \cite{Lenz:2011ti}
\begin{eqnarray}
\phi_d^{\text{SM}} = (-7.5 \pm 2.4) \times 10^{-2}~,\ \phi_s^{\text{SM}} = ( 3.8 \pm 1.1) \times 10^{-3}~.
\end{eqnarray}

The flavor specific charge asymmetry $a_{s\ell}^q$ is related to the mass and width differences in the $B_q - \overline{B}_q$ system as
\begin{equation}
  a_{s\ell}^q =\mbox{Im} \frac{\Gamma_{12}^q}{M_{12}^q} = \frac{|\Gamma_{12}^q|}{|M_{12}^q|} \sin \phi_q = \frac{\Delta \Gamma_q}{\Delta M_q} \tan \phi_q\,. \label{eq:asanal}
\end{equation}
Here, the experimental value of $\Delta M_s$ obtained from the LHCb 0.34fb$^{-1}$ with 68.3\% C.L. is \cite{LHCb-CONF-2011-050}
\dis{
\Delta M_s &=17.725\pm 0.041(\rm stat.)\pm 0.026(\rm sys.)~ {\rm ps}^{-1}~,  \label{eq:DMexp}
}
(the combined result of CDF and D0 is $\Delta M_s =17.78\pm 0.12~{\rm ps}^{-1}$)
while the SM prediction is \cite{Barberio:2008fa}
\begin{eqnarray}
(\Delta M_s)^{\rm SM} = (17.3 \pm 2.6) \, \text{ps}^{-1}
\label{eq:delmssm}
\end{eqnarray}
which corresponds to $f_{B_s} = $231 MeV and $\hat B_B$ = 1.28 of Eqs.\eqref{eq:B-hat} and \eqref{eq:<O^VLL>} \cite{Lenz:2011ti}.

The observed value of $\Delta M_s$ has not so much deviated from the SM prediction. 
Without considering the NP contribution to $\Gamma^s_{12}$,  therefore, it is impossible to obtain the observed central value of $a_{s\ell}^s$ from Eqs. (\ref{eq:asanal}) and (\ref{eq:DMexp}) for $q=s$ even we assume $\sin\phi_s = -1$. With the recent LHCb bound for $\phi^{J/\psi\,\phi}_s$, the maximally possible enhancement of $a_{s\ell}^s$ in this case is outside the boundary of $1\sigma$ of the observed value in (\ref{eq:base}) as seen in Fig. \ref{fig:tauhs} (b).
Therefore, an additional NP contribution to $\Gamma_{12}^s$ is preferred to explain the like-sign dimuon charge asymmetry through the $B_s - \bar{B}_s$ mixing.

To probe the NP contribution, we split $\Gamma_{12}$ or $M_{12}$ to the SM and NP contributions as
\begin{eqnarray}
\frac{\Gamma_{12}^{q\, \rm NP}}{\Gamma_{12}^{q\, \rm SM}} \equiv \tilde{h}_q e^{i 2 \tilde{\sigma}_q}~,~
\frac{M_{12}^{q\, \rm NP}}{M_{12}^{q\, \rm SM}} \equiv h_q e^{i 2\sigma_q}~,
\label{eq:hs_and_sigmas}
\end{eqnarray}
for real and non-negative parameters $\tilde{h}_q$ and $h_q$, with the phases constrained in the region, $0 \le \sigma_q, \tilde{\sigma}_q \le \pi$. Then, the flavor specific charge asymmetry is given by \cite{Kim:2010gx}
\dis{ 
a_{s\ell}^q &= \frac{|\Gamma_{12}^{q\, \rm SM}|}{|M_{12}^{q\, \rm SM}|} \frac{1}{1+h_q^2+2h_q\cos2\sigma_q} \\
&\hspace{1.7cm}\times \left[ \left\{-\tilde{h}_q \sin2\tilde{\sigma}_q (1+h_q\cos2\sigma_q) + h_q \sin2\sigma_q (1+\tilde{h}_q\cos2\tilde{\sigma}_q)\right\} \cos\phi_q^{\rm SM} \right. \\
& \hspace{2.5cm} + \left. \left\{ (1+\tilde{h}_q \cos2\tilde{\sigma}_q)(1+h_q \cos2\sigma_q) + h_q \tilde{h}_q \sin2\sigma_q\sin2\tilde{\sigma}_q \right\} \sin\phi_q^{\rm SM} \,\,  \right]~.
\label{eq:mastereq}
} 
Also, the ratio of Eq.\eqref{eq:mastereq} to its SM value is given by
\dis{
-a_{s\ell}^q/(a_{s\ell}^q)^{\rm SM} &= \frac{1}{1+h_q^2+2h_q\cos2\sigma_q} \\
&\hspace{0.7cm}\times \left[ \left\{\tilde{h}_q \sin2\tilde{\sigma}_q (1+h_q\cos2\sigma_q) - h_q \sin2\sigma_q (1+\tilde{h}_q\cos2\tilde{\sigma}_q)\right\} \cot\phi_q^{\rm SM} \right. \\
& \hspace{1.5cm} - \left. \left\{ (1+\tilde{h}_q \cos2\tilde{\sigma}_q)(1+h_q \cos2\sigma_q) + h_q \tilde{h}_q \sin2\sigma_q\sin2\tilde{\sigma}_q \right\} \,\,  \right]~.
\label{eq:ra}
}
Note that the factor $1/(1+h_q^2+2h_q\cos2\sigma_q)$  in Eq. \eqref{eq:ra} is fixed by the ratio of $\Delta M_q^{\rm SM} / \Delta M_q$, near to 1. Therefore a sizable NP contribution to $|\Gamma_{12}^{q\,{\rm NP}}/\Gamma_{12}^{q\,{\rm SM}}| = \tilde{h}_q$ is necessary if can take the dominant role in explaining the observed dimuon charge asymmetry. 

The $Z'$ models to enhance the $a_{s\ell}^s$ require the existence of nonzero off-diagonal couplings $g_{sb}^L$ and $g_{sb}^R$, where $g_{\psi \chi}^{L,R}$ is the coupling of $Z'$ to fermions $\psi_{L,R}$ and $\chi_{L,R}$. Turning off one of the couplings $g_{sb}^L$ and $g_{sb}^R$ for simplicity, this scenario demands the existence of rather large couplings $|g_{\tau \tau}^{L,R}| > 1$ to explain the asymmetry within $1\sigma$ range from the observed central value due to the strict $\Delta M_s$ constraint. The situation is the same even in the case that the mass of $Z'$ is similar to that of the $Z$ boson.
Such large $g_{\tau \tau}^{L,R}$ couplings can violate the observations in the electroweak precision test (EWPT).
Therefore, we need to turn on both of the flavor changing couplings $g_{sb}^L$ and $g_{sb}^R$. The scenario considering the $g_{\tau \tau}^{L,R}$ couplings to explain the dimuon charge asymmetry will be called as  ``{\it $g_{\tau \tau}$ scenario}" in this paper.  

On the other hand, considering the nonzero $Z'$ coupling to the charm quark pair can also explain the dimuon charge asymmetry by considering the interference of the NP contribution and the SM process. Due to the interference, the couplings $g_{sb}^{L,R} g_{cc}^{L,R}$ contribute to $a_{s\ell}^s$ linearly while $g_{s,b}^{L,R} g_{\tau \tau}^{L,R}$ do quadratically so that the interference effect dominates the enhancement of $a_{s\ell}^s$ unless the NP contribution is larger than that of the SM. Therefore, it is possible to explain the asymmetry with rather smaller $Z'$ couplings in this scenario so that we can avoid the direct constraint such as the decay of $B_s \to D D_s$ \cite{Alok:2010ij}. The scenario considering such contribution will be called as ``{\it $g_{cc}$ scenario}" in this paper. Describing the corresponding $\Gamma_{12}^s$ in each of our $Z'$ scenario, there are six real free  parameters, \ie the complex $g_{sb}^{L,R}$ and the real $g_{\tau \tau}^{L,R}$ ($g_{cc}^{L,R}$) since the diagonal couplings have to be real.

Every experimental result depends not only on the mass of $Z'$ but also on its couplings to the matter because the new interaction depends on the ratio $(g_{\psi \chi}^{L,R}/g_1) (M_Z / M_{Z'})$, where $g_1 = g/\cos\theta_W$ for $g$ is the SU(2)$_L$ coupling and $\theta_W$ is the weak mixing angle. Therefore, the experimental bounds can be applied for any values of $M_{Z'}$ by proper rescaling of the couplings $g_{\psi \chi}^{L,R}$. According to this fact, we set the reference value $M_{Z'} = M_Z$ for the representation of our analyses so that one can easily see the results for any $M_{Z'}$ one wants to analyze, by simple rescaling of the $Z'$ couplings. Actually, our reference value of $M_{Z'}$ is not unrealistic since the $b$-quark forward-backward asymmetry $A_{FB}^b$ at the LEP can be explained in terms of $Z'$ where $M_{Z'} \approx M_Z$ and the non-zero $g_{ee}^{L,R}$ and $g_{bb}^{L,R}$ exist \cite{Dermisek:2011xu}. As a conservative approach, one can consider the heavy $Z'$ whose mass is much larger than 1 TeV to avoid the current experimental limits when the $Z'$ couplings to matter are SM-like \cite{Acosta:2005ij,Chatrchyan:2012hd}. By simply rescaling our final result in such a case, some $Z'$ couplings to the matter should be much larger than 1 to explain the dimuon charge asymmetry, which is unrealistic in the perturbative regime. On the other hand, one can also consider very light $Z'$ cases whose couplings are small enough to avoid the direct $Z'$ search bounds. Then, one needs to apply the other experimental bounds which we will explain from now on.

The NP models accommodating the sizable new contribution in $\Gamma_{12}^q$ suffer from the various experimental bounds, mainly due to the recently updated LHCb data of 1fb$^{-1}$. In the next section, we analyze the related bounds in detail by focussing on the enhancement of $a_{s\ell}^s$.


\section{Experimental constraints}
\label{sec:ex}

In this section, we analyze the various experimental constraints in obtaining the new sizable contribution to $\Gamma_{12}^s$ from the $B_s -\bar{B}_s$ mixing. The NP contribution to $\Gamma_{12}^s$ via the operator $(\bar{s} b)(\bar{f} f)$ where $f$ is a SM fermion can also affect the various $B_s$ or $B$ meson decay processes {\footnote{$B$ generically denotes $B_d^0, B_d^{\pm}$ mesons.}}. 
In the $Z'$ models, the new contribution is realized by the tree level FCNC process, which can be large enough to threaten the current experimental bounds. In this section, we introduce the experimental constraints from $\Delta M_s$, $\Delta \Gamma_s$, $\phi_s^{J/\psi\,\phi}$, $b \to s \nu \bar{\nu}$, and $B \to J/\psi K_S$. Then, we will show what extent the NP parameter space explaining the dimuon charge asymmetry can be constrained by such bounds, by applying our $Z'$ scenarios. 

For the simplicity, we turn off the couplings $g_{\ell \ell}^{L,R}$ for the light leptons $\ell = e^-, \mu^-$ not to consider the tree level NP contribution in the observations such as $B \to X_s \ell^+ \ell^-$, $B \to K^{\ast} \ell^+ \ell^-$, and $B_s \to \ell^+ \ell^-$ as shown in \cite{Kim:2010gx}. For the case $g_{bb}^{L,R} \ne 0$, a one-loop induced NP contribution can affect the $b \to s \gamma$. Such contribution is well summarized in our Appendix \ref{appen:bsg} for the future use.

\subsection{$\Delta M_s$}
\label{sec:delms}

\begin{figure}
\includegraphics[width=9cm]{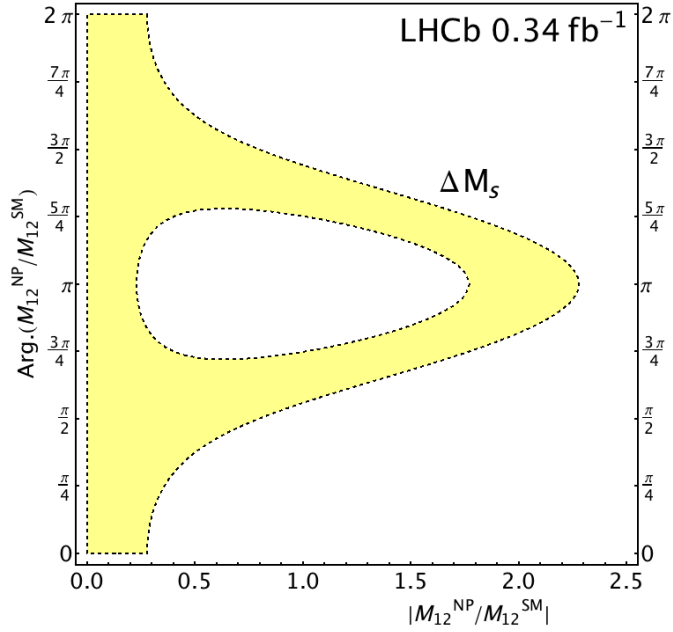}
\caption{The yellow colored region denotes the parameter space allowed by the 90\% C.L. ($1.65\sigma$) experimental bounds of $\Delta M_s$ observed at the LHCb 0.34fb$^{-1}$. The rough upper limit of $h_s$ is 2.3 according to this figure. The results from the CDF and D0 are not so much different from this.}
\label{fig:delms}
\end{figure}

The experimental measurements of $\Delta M_s$ both from the LHCb and the Tevatron have no significant deviation from the SM prediction. Therefore, the allowed parameter space is highly constrained as shown in Fig. \ref{fig:delms} in terms of the general parameters $h_s$ and $2\sigma_s$.

In the SM, the dominant contributions to $M_{12}$ come from the top quark loops and their effects are summarized as follow.
\begin{eqnarray}
M_{12}^{\rm SM}=\frac{G_F^2}{12\pi^2} M_W^2 (V_{tb} V_{ts}^{\ast})^2 S_0(\bar m_t^2/M_W^2)  m_{B_s} f^2_{B_s} \eta_{2B}   \hat B_B
\label{eq:M_12^SM}
\end{eqnarray}
Here, $S_0(x)$ 
is an Inami-Lim function for the corresponding box diagrams \cite{Inami:1980fz}, and $\eta_{2B} $ and $\hat B_B$ are $\mu_b$ and $\mu_W$ independent quantities at a given order of QCD corrections.
At the NLO, $\eta_{2B} \simeq 0.551$ \cite{Buchalla:1995vs} and $\hat B_B$ is given as 
\begin{eqnarray}
\hat B_B = [ \alpha_s (\mu_b) ]^{-6/23} \left(1+ \frac{\alpha_s (\mu_b)}{4\pi} J_5 \right) B_1^{VLL} (\mu_b)~,
\label{eq:B-hat}
\end{eqnarray} 
where $J_5 = 1.627$ (in NDR and $f=5$) \cite{Buchalla:1995vs}. $B_1^{VLL}$ is a bag parameter of a matrix element
\begin{eqnarray}
 \langle B_s | O^{VLL}_1(\mu) | \bar{B}_s \rangle &=& \frac{2}{3} m_{B_s}^2 f_{B_s}^2 B_1^{VLL} (\mu)~, 
\label{eq:<O^VLL>}
\end{eqnarray}
where
\begin{eqnarray}
O^{VLL}_1 &=&  (\bar s_L \gamma ^\mu b_L) (\bar s_L \gamma _\mu b_L)
\label{eq:}
\end{eqnarray}
and $m_{B_s}$ and $f_{B_s}$ are ${B_s}$ meson mass and its decay constant, respectively.

For the evaluation of Eq.\eqref{eq:M_12^SM}, we use 
$G_F = 1.16637(1) \times 10^{-5}$ GeV$^{-2}$,
$M_W = 80.399(23)$ GeV,
$m_{B_s} = (5366.3 \pm 0.6)$ MeV,
$|V_{tb}| = 0.999152^{+0.000030}_{-0.000045}$,
$|V_{ts}| = (4.03^{+0.11}_{-0.07})$
\cite{PDG2011}.
\footnote{Here and after, the figures in parentheses after the values give the 1-standard-deviation uncertainties
in the last digits} For the top-quark mass, we use $m_t^{pole} = 173.2 \pm 0.9$ GeV (correspondingly $\bar m_t (\bar m_t) = 165.8 \pm 0.9$ GeV )  \cite{Lancaster:2011wr}. 
Finally we use $f_{B_s} = (229 \pm 6)$ MeV and  $\hat B_B = 1.291 \pm 0.043$ \cite{Lenz:2012az}.

For the Z$'$ and its flavor violating interactions of $g_{sb}^L \bar s_{L} \gamma^\mu b_{L} Z'_\mu$ and $g_{sb}^R \bar s_{R} \gamma^\mu b_{R} Z'_\mu$, following effective operators, in addition to $O^{VLL}_1$, are induced at the scale where the $Z'$ is integrated out.
\begin{eqnarray}
O^{VRR}_1 =  (\bar s_R \gamma ^\mu b_R) (\bar s_R \gamma _\mu b_R) \\
O^{LR}_1 =  (\bar s_L \gamma ^\mu b_L) (\bar s_R \gamma _\mu b_R) 
\label{eq:O^LR_1} \\
O^{RL}_1 =  (\bar s_R \gamma ^\mu b_R) (\bar s_L \gamma _\mu b_L)
\label{eq:O^RL_1}
\end{eqnarray}
At the same time,  QCD corrections to the operators of Eqs. \eqref{eq:O^LR_1} and \eqref{eq:O^RL_1} induce following operators as well.
\begin{eqnarray}
O^{LR}_2 =  (\bar s_R b_L) (\bar s_L b_R) \\
O^{RL}_2 =  (\bar s_L b_R) (\bar s_R b_L) 
\end{eqnarray}
Using those notations, we write down an effective Hamiltonian at the scale $\mu_b$ as
\begin{eqnarray}
{\cal H}_{eff}^{Z'} 
&=& \frac{1}{2M_{Z'}^2} \Big[ \, \eta^{LL}  \left( (g_{sb}^L)^2   + (g_{sb}^R)^2  \right)  \mathcal{O}^{VLL}_1 
+ \, 2\, \eta_{11}^{LR} \, g_{sb}^L g_{sb}^R \, \mathcal{O}_1^{LR} 
+ 2\, \eta_{21}^{LR} g_{sb}^L g_{sb}^R \, \mathcal{{O}}_2^{LR} \Big] 
\label{eq:}
\end{eqnarray} 
Note that we identify $ \mathcal{O}_i^{XY} $ with $ \mathcal{O}_i^{YX} $ ($X, Y = L \text{ or } R $) at this stage, reflecting the fact that the QCD is vector-like and corresponding matrix elements are equal. The QCD corrections are given at the NLO in Ref. \cite{Buras:2001ra} as
\begin{eqnarray}
\eta^{LL}
&=&  \eta_5^{6/23}+ \frac{\alpha_s (m_b)}{4\pi} \left(1.63\left(1-\eta_5\right) \eta_5^{6/23}\right) \nonumber  \\
\eta_{11}^{LR}
&=& \eta_5^{3/23} + \frac{\alpha_s (m_b)}{4\pi} \left(0.93 \eta_5^{-24/23} + \eta_5^{3/23} \left( -2.10 + 1.17 \eta_5 \right)  \right) \nonumber \\
\eta_{21}^{LR}
&= &  \frac23 \left(\eta_5^{3/23} - \eta_5^{-24/23}\right) 
+ \frac{\alpha_s (m_b)}{4\pi}\left((-11.73+0.78\eta_5) \eta_5^{3/23} + \eta_5^{-24/23} \left( -5.30 + 16.25 \eta_5 \right) \right) \nonumber 
\label{eq:}
\end{eqnarray} 
where $\eta_5 \equiv \alpha^{(5)}_s (\mu_Z)/\alpha^{(5)}_s (\mu_b)$. 
Parametrizing the hadronic matrix elements as
\begin{eqnarray}
 \langle B_s | \mathcal{O}^{LR}_1(\mu) | \bar{B}_s \rangle  &=& -\frac13 \left( \frac{m_{B_s}}{\bar m_b + \bar m_s} \right)^2 m_{B_s}^2 f_{B_s}^2 B_1^{LR} (\mu)~, \\
\langle B_s | \mathcal{O}^{LR}_2 (\mu)| \bar{B}_s \rangle  &=& \frac12 \left( \frac{m_{B_s}}{\bar m_b + \bar m_s} \right)^2 m_{B_s}^2 f_{B_s}^2 B_2^{LR} (\mu)~, 
\end{eqnarray}
we get following expression for the $Z'$ contrition to $M_{12}$.
\begin{eqnarray}
M_{12}^{Z'} 
&=&  \frac{ m_{B_s} f_{B_s}^2}{6 M_{Z'}^2} \left[ \, \eta^{LL} \left( (g_{sb}^L)^2  + (g_{sb}^R)^2 \right) B^{VLL}_1
-  g_{sb}^L g_{sb}^R  \cdot  \left( \frac{m_{B_s}}{\bar m_b + \bar m_s} \right)^2  \left( \eta^{LR}_{11} B_1^{LR}  
- \frac{3}{2}   \eta^{LR}_{21}   B_2^{LR} \right) \right]~. \nonumber \\
\label{eq:M_12^Z'}
\end{eqnarray}
For the evaluation of Eq.\eqref{eq:M_12^Z'}, we use the two-loop RG running with the input of $\alpha_s (M_Z) = 0.1184(7)$, $\alpha_s(\mu_b)$ is evaluated at 4.6 GeV where the bag parameters are provided as $B^{VLL}_1 (m_b) = 0.87 \pm 0.05$, $B_1^{LR} (m_b) = 1.75 \pm 0.21$, and $B_2^{LR} (m_b) = 1.16 \pm 0.07$ \cite{Becirevic:2001xt}. ($\bar{m}_s (2\text{GeV}) = 100^{+30}_{-20}$ MeV \cite{PDG2011} is evaluated as $\bar m_s$(4.6 GeV) = $83^{+25}_{-17}$ MeV.) With these inputs, we obtain 
\begin{eqnarray}
\quad h_s = (7.53 \times 10^5) \cdot \Big| \left( g_{sb}^L \right)^2 + \left( g_{sb}^R \right)^2 - k \cdot g_{sb}^L g_{sb}^R \, \Big|
\label{eq:hsgeneral}
\end{eqnarray}
where $k = 5.05 \pm 0.47$.

The value of $h_s$ should be as small as $< 2.3$ to satisfy the experimental constraint of (\ref{eq:DMexp}). Therefore, the terms inside the squared bracket of (\ref{eq:hsgeneral}) must be as small as $\lesssim 3 \times 10^{-5}$. This result can be rewritten as   
\dis{
\Big| \left( g_{sb}^L \right)^2 + \left( g_{sb}^R \right)^2 - k \cdot g_{sb}^L g_{sb}^R \, \Big| \lesssim 3.06 \times 10^{-6}~.
\label{eq:finetune}
} 
Eq.\eqref{eq:finetune} describes a complex hyperbolic surface which is flipped along the asymptotic complex lines satisfying 
\dis{
\left( g_{sb}^L \right)^2 + \left(g_{sb}^R \right)^2-k\cdot  g_{sb}^L g_{sb}^R  = 0~ \label{eq:asym}~,
} 
or equivalently,
\dis{
g_{sb}^R = a g_{sb}^L~, \hspace{0.5cm} g_{sb}^R = (1/a) g_{sb}^L~,
\label{eq:asymline}
}
where $a = 4.84$ for $k=5.05$. On these asymptotic lines, $\theta_L = \theta_R$ where $\theta_{L,R}$ is the phase of $g_{sb}^{L,R}$ respectively. Consequently, the bound (\ref{eq:finetune}) indicate that the generic values of $|g_{sb}^{L,R}|$ must be smaller than $10^{-3}$ unless they are within (or close to) the asymptotic lines (\ref{eq:asymline}). Since the $\Delta M_s$ constraint parametrized by (\ref{eq:finetune}) is highly stringent for $|g_{sb}^{L,R}| > 10^{-3}$, the parameter space containing such values of couplings cannot avoid the fine tuning. 

For the case that one of $g_{sb}^{L,R}$ is turned off, we can easily induce that the absolute value of the remaining nonzero coupling must be definitely smaller than $1.75 \times 10^{-3}$. Therefore, the required value of $|g_{\tau \tau}^{L,R}|$ or $|g_{cc}^{L,R}|$ for the explanation of the dimuon charge asymmetry in this case must be larger than 1, which is easily induced from analyzing the results in \cite{Kim:2010gx,Alok:2010ij}.


\subsection{$\Delta \Gamma_s$ and $\phi_s^{J/\psi\,\phi}$ from $B_s \to J/\psi\,\phi$}
\label{sec:lhcb}

The enhancement of the like-sign dimuon charge asymmetry is constrained by the experimental measurement of the width difference $\Delta \Gamma_s$ of the mass eigenstate $B_s^0$ mesons, and the phase difference $\phi_s^{J/\psi\,\phi}$ between the $B_s$ mixing and the $b \to s c \bar{c}$ decay. These are simultaneously determined by measuring the indirect CP asymmetry of $B_s \to J/\psi\,\phi$ decay. The recent result from the LHCb of 1fb$^{-1}$ integrated luminosity shows that \cite{lhcbn1}
\begin{eqnarray}
\Delta \Gamma_s &=& 0.116 \pm 0.018 ({\rm stat.}) \pm 0.006 ({\rm syst.})~ \text{ps}^{-1}~~,
\label{eq:deltagam} \\
\phi_s^{J/\psi\,\phi} &=&  -0.001\pm 0.101 ({\rm stat.}) \pm 0.027 ({\rm syst.})~\text{rad}~~, \label{eq:spp}
\end{eqnarray}
in which $\Delta \Gamma_s$ has about 1.2$\sigma$ deviation{\footnote{Note that the sign of $\Delta \Gamma_s$ is fixed to be positive in this result.}} from $(\Delta \Gamma_s)^{\rm SM} = (0.087 \pm 0.021)$ ps$^{-1}$ and $\phi_s^{J/\psi\,\phi}$ agrees well with the SM prediction $(\phi_s^{J/\psi\,\phi})_{\rm SM} =$ 
Arg. ($(V_{ts}V_{tb}^{\ast})^2/(V_{cs} V_{cb}^{\ast})^2)=-2\beta_s^{\rm SM} = -0.036 \pm 0.002$ \cite{Lenz:2011ti}. Such new LHCb results dramatically reduce the room of new physics contribution in $B_s - \bar{B}_s$ mixing compared to those of the previous LHCb (337 pb$^{-1}$), the CDF (5.2 fb$^{-1}$), and the D0 (8.0 fb$^{-1}$).

We first deal with the issue related with $\phi_s^{J/\psi\,\phi}$, whose measurement at the LHCb 1fb$^{-1}$ shows the most dramatic changes compared to the previous ones. The analytic expression of $\phi_s^{J/\psi\,\phi}$ is well summarized in \cite{Chiang:2009ev} and \cite{Alok:2010ij}. Neglecting the SM strong phases in the $B_s \to J/\psi\,\phi$ process, we obtain \cite{Chiang:2009ev,Alok:2010ij}
\begin{eqnarray}
\sin\phi_s^{J/\psi\,\phi} =\sin (-2\beta_s + \phi_M^s) + 2 |r_\lambda| \cos (-2\beta_s + \phi_M^s) \sin \varphi_\lambda~,
\label{eq:phisjp}
\end{eqnarray} 
where $\phi_M^s=$ Arg.$(M_{12}/M_{12}^{\rm SM})$ is from the NP contribution in the dispersive part of $B_s -\bar{B}_s$ mixing, and the term with $r_\lambda$ is from the NP contribution in the $b \to sc\bar{c}$ decay. Note that this result is obtained using the approximation that $|r_\lambda| \ll 1$ from the exact relation in \cite{Chiang:2009ev}. In the figures to show the allowed parameter space, we use the exact relation. 

\begin{figure}
\subfigure[\ Experimental bounds in the $B_s$ sysmtem]{
\includegraphics[width=7.5cm]{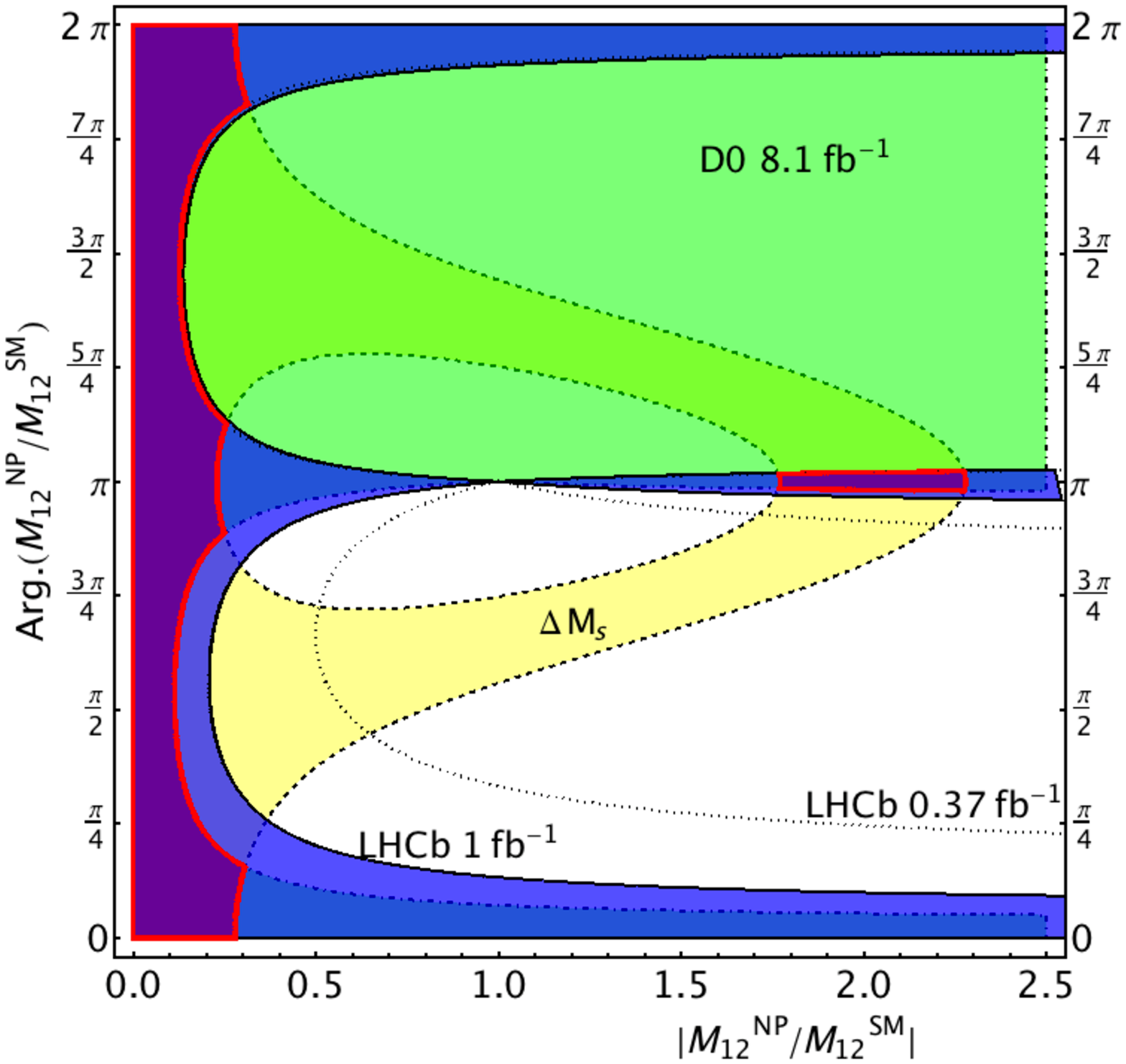}
}
\quad
\subfigure[\ Possible enhancement if $\Gamma_{12}^s = \Gamma_{12}^{s\,{\rm SM}}$]{
\includegraphics[width=7.5cm]{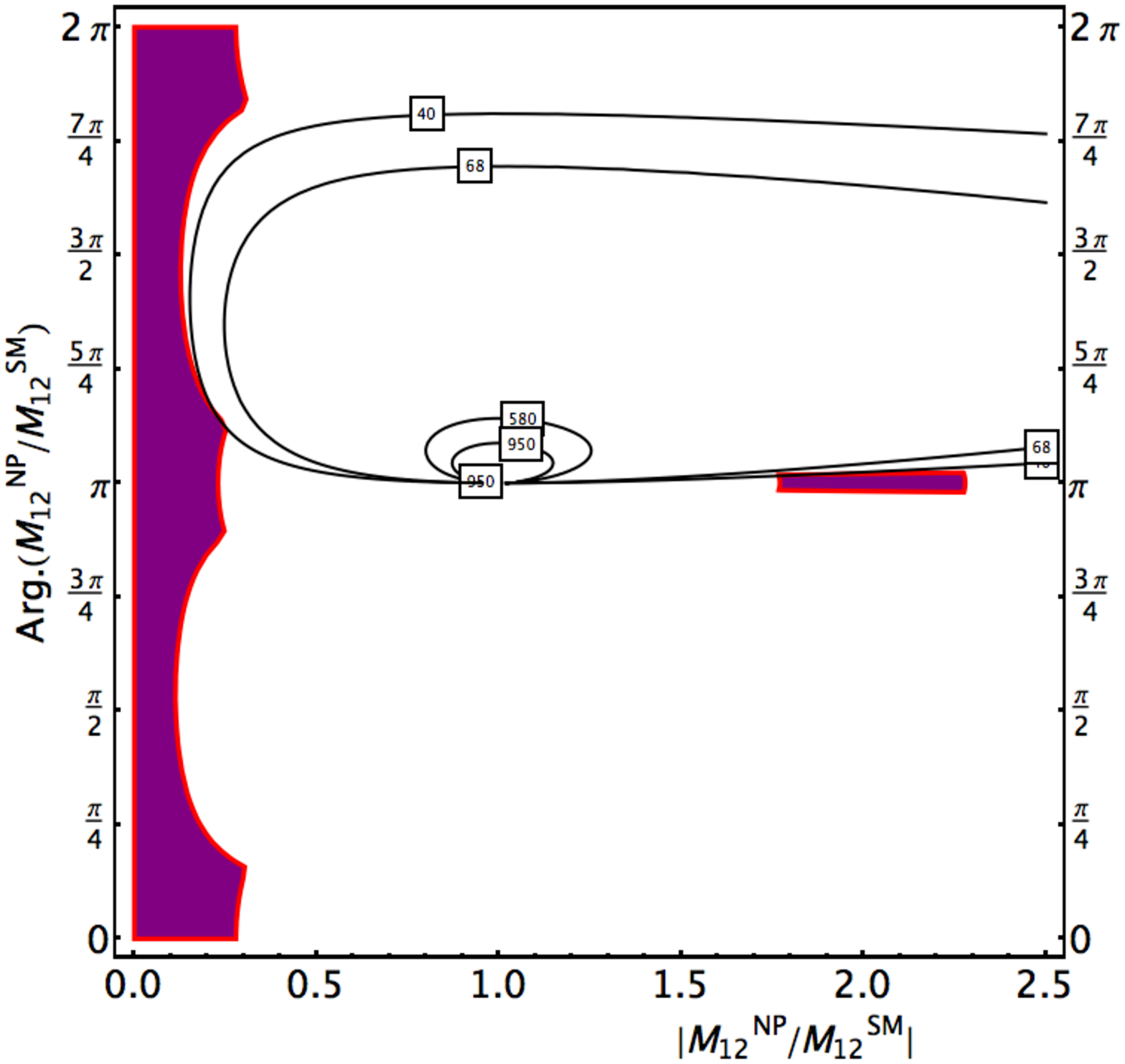}
}
\caption{The experimental bounds apply for the general NP scenarios {\it without new phases in $b \to s c\bar{c}$}, such as the {\it $g_{\tau \tau}$ scenario}. (a) The final allowed region of $h_s = |M_{12}^{\rm NP}/M_{12}^{\rm SM}|$ and $2\sigma_s =$ Arg.($M_{12}^{\rm NP} / M_{12}^{\rm SM}$) from the $\Delta M_s$, $\phi_s^{J/\psi\,\phi}$ (LHCb 1fb$^{-1}$, 0.37fb$^{-1}$, and D0 8fb$^{-1}$) constraints is shown as the purple color surrounded by the thick red line. The yellow region (inside the dashed line boundary) : allowed by 90\% $\Delta M_s$.  The green region (inside the dot-dashed line boundary) : allowed by 90\% $\phi_s^{J/\psi\,\phi}$ at the 8.0 fb$^{-1}$ D0. ($\phi_s^{J/\psi\,\phi} = 0.15 \pm 0.18({\rm stat.}) \pm 0.06 ({\rm syst.})$ \cite{d0phi}). The blue region (inside the line boundary) : allowed by 90\% $\phi_s^{J/\psi\,\phi}$ at the recent 1.0 fb$^{-1}$ LHCb and the boundary at the 0.37 fb$^{-1}$ at the LHCb in the last year is denoted as the dotted lines.  The purple region surrounded by the thick red line denotes the allowed parameter space from all the commented constraints. The mainly remained region is $h_s < 0.3 \ll 1$, which provides a fine tuning choice in the parameter space explaining $a_{s\ell}^s$. The other region of $2\sigma_s \sim \pi$ with $1.7 < h_s < 2.2$ is irrelevant in the enhancement of $a_{s\ell}^s =$ Im$(\Gamma_{12}^s/M_{12}^s)$ in the {\it $g_{\tau \tau}$ scenario}. 
(b) Without NP contribution to $\Gamma_{12}$, we represent the possible enhancement of $-a_{s\ell}^s/(a_{s\ell}^s)^{\rm SM}$ with the numbers and the contours. In this case, the enhancement is quite limited such that $-a_{s\ell}^s/(a_{s\ell}^s)^{\rm SM} < 40$ to be consistent with all the experimental bounds.}
\label{fig:tauhs}
\end{figure}

When there is no NP phase contribution in $b \to s c\bar{c}$ process like our {\it $g_{\tau \tau}$ scenario} in $Z'$, we have $r_\lambda=0$. Then, the NP effect in $\phi_s^{J/\psi\,\phi}$ contributes only through $\phi_M^s$.
Since we know that $\sin \phi_M^s = h_s \sin 2\sigma_s / \sqrt{1+h_s^2 + 2h_s \cos2\sigma_s} = (h_s \sin2\sigma_s)\sqrt{\frac{\Delta M_s^{\rm SM}}{\Delta M_s}} \approx h_s \sin2\sigma_s$ by combining the experimental bound of $\Delta M_s$, the value of $h_s \sin2\sigma_s$ must be small enough to satisfy the measured result (\ref{eq:spp}). In Fig. \ref{fig:tauhs} (a), this result is represented in terms of our parameters $2\sigma_s$ and $h_s$ with the purple color surrounded by the thick red line. To satisfy both of the experimental bounds $\Delta M_s$ (\ref{eq:DMexp}) and $\phi_s^{J/\psi\,\phi}$ (\ref{eq:spp}), the allowed parameter space must satisfy $h_s < 0.3 \ll 1$ except the small region around $2\sigma_s \sim \pi$ with $1.7 < h_s < 2.2$. 

Without NP contribution to $\Gamma_{12}$, we can see that the enhancement in $a_{s\ell}^s$ is quite limited as 
$-a_{s\ell}^s/(a_{s\ell}^s)^{\rm SM} < 40$ in Fig. \ref{fig:tauhs} (b), which is not satisfactory to explain the dimuon charge asymmetry within $1\sigma$. Considering the coefficient $\cot\phi_s^{\rm SM}$ of $\tilde{h}_s$ in (\ref{eq:ra}), we hence need at least $\mathcal{O}(0.1)$ contribution by $|\Gamma_{12}^{\rm NP}/\Gamma_{12}^{\rm SM}|$ to $a_{s\ell}^s$. Applying our later result (\ref{eq:thstau}) in case of the {\it $g_{\tau \tau}$ scenario}, the value of coupling $|g_{sb}^{L,R} g_{\tau \tau}^{L,R}|$ should be much larger than $10^{-3}$. Restricting $|g_{\tau \tau}^{L,R}|<1$ to avoid the nonperturbativity bound, we need $|g_{sb}^{L,R}| \gg 10^{-3}$ which must be around the asymptotic lines (\ref{eq:asymline}) to satisfy the $\Delta M_s$ bound. Therefore, we can generically set $\theta_L = \theta_R$ in the {\it $g_{\tau \tau}$ scenario}, which induces $2\tilde{\sigma_s} = 2\sigma_s + n\pi$ for an integer $n$ neglecting the contribution by the SM phases. We can conclude from this, at the region around $2\sigma_s \sim \pi$ with $1.7 < h_s < 2.2$, the enhancement in $a_{s\ell}^s =$ Im$(\Gamma_{12}^s/M_{12}^s)$ is ignorable in our {\it $g_{\tau \tau}$ scenario}. In result, we will proceed the analysis with the condition $h_s < 0.3$ in this scenario.

On the other hand, for the cases that we have NP phase contribution in $b \to s c\bar{c}$ process like our {\it $g_{cc}$ scenario}, the NP effect in $\phi_s^{J/\psi\,\phi}$ contributes also through $r_\lambda \ne 0$. The NP contribution in the $B_s \to J/\psi\,\phi$ amplitude is parametrized as
\begin{eqnarray}
\sum \langle (J/\psi\,\phi)_\lambda | \mathcal{O}_{\rm NP} | B_s \rangle = b_\lambda e^{i\varphi_\lambda} ~,
\end{eqnarray}
where $\lambda$ is the polarization of final state vector particles. The longitudinal direction is $\lambda =0$ and the two transverse directions are $\lambda =\{ + , -\}$. The angle $\phi_\lambda$ is the new weak phase from the above NP contribution. The ratio of the amplitude $|r_\lambda| = b_\lambda / a_\lambda$ is defined for the SM amplitude $a_\lambda$.

In the {\it $g_{cc}$ scenario} of $Z'$ model, we obtain the following result according to \cite{Chiang:2009ev} such that
\begin{eqnarray}
|r_{\lambda=0}| &=& \left| \frac{1}{g_1^2} \frac{M_Z^2}{M_{Z'}^2} \frac{2(g_{cc}^L + g_{cc}^R)(g_{sb}^L - k_0 g_{sb}^R)}{V_{cb} V_{cs}^{\ast} \cdot 0.17} \right|~, \\
|r_{\lambda=+}| &=& \left| \frac{1}{g_1^2} \frac{M_Z^2}{M_{Z'}^2} \frac{2(g_{cc}^L + g_{cc}^R)(g_{sb}^L - k_+ g_{sb}^R)}{V_{cb} V_{cs}^{\ast} \cdot 0.17} \right|~, \\
|r_{\lambda=-}| &=& \left| \frac{1}{g_1^2} \frac{M_Z^2}{M_{Z'}^2} \frac{2(g_{cc}^L + g_{cc}^R)(g_{sb}^L - k_- g_{sb}^R)}{V_{cb} V_{cs}^{\ast} \cdot 0.17} \right|~,
\end{eqnarray}
where $k_0 = 1$, $k_+ = 8.8, 9.8$ and $k_-=0.11, 0.10$ depending on the model of the form factors Melikhov-Stech \cite{Melikhov:2000yu} and Ball-Zwicky \cite{Ball:2004rg}, respectively.{\footnote{Actually, there are typically about 10 $\%$ theoretical uncertainties in the form factors. Such consideration in $k_+$ as an example is shown in our Appendix \ref{appen:form}.}} The vector interaction of the charm quark pair is obtained from the factorization $\langle J/\psi | \bar{c} \gamma^{\mu} c | 0 \rangle$. 

Consequently, we obtain the following expression in the {\it $g_{cc}$ scenario} neglecting the SM prediction for $|r_\lambda| < 1$. 
\dis{
\sin \phi_s^{J/\psi\,\phi} &= \frac{ h_s \sin 2\sigma_s}{\sqrt{1+h_s^2 + 2h_s \cos2\sigma_s}} \\
&\hspace{1cm} + \left| \frac{2}{g_1^2} \frac{M_Z^2}{M_{Z'}^2} \frac{2 (g_{cc}^L + g_{cc}^R) (g_{sb}^L - k_{0,\pm} g_{sb}^R)}{V_{cb} V_{cs}^{\ast} \cdot 0.17} \right| \frac{1+ h_s \cos 2\sigma_s}{\sqrt{1+h_s^2 + 2h_s \cos2\sigma}} (\sin\varphi_{0,\pm})~. 
\label{eq:jppanal}
}

If we simply assume $h_s \approx 0$ which is conservatively safe from the $\Delta M_s$ bound, we can neglect the first term of Eq. (\ref{eq:jppanal}) and the expression is simplified as
\dis{
\sin \phi_s^{J/\psi\,\phi} \approx  (1.0 \times 10^3) \left|(g_{cc}^L + g_{cc}^R) (g_{sb}^L - k_{0,\pm} g_{sb}^R)\right| (\sin\varphi_{0,\pm})~. 
}
To satisfy the recent LHCb result of 1fb$^{-1}$ with 90\% C.L., we obtain the following simple condition on the couplings in this case 
\dis{
-1.7 \times 10^{-4} < \left|(g_{cc}^L + g_{cc}^R) (g_{sb}^L - k_{0,\pm} g_{sb}^R)\right| (\sin\varphi_{0,\pm}) < 1.7 \times 10^{-4}~,
\label{eq:jpex}
}
which provides a strong constraint on the values of $|g_{sb}^{L,R} g_{cc}^{L,R}|$ unless the $Z'$ vector coupling to the charm quark pair is axial. For $|g_{sb}^L| \ll k_+ |g_{sb}^R|$,  the most stringent bound is obtained from the $\lambda=+$ case and $|\sin \varphi_+| \approx |\sin\theta_R|$.  For the other case, the most stringent bound is obtained from the $\lambda=-$ case and $|\sin \varphi_-| \approx |\sin\theta_L|$. Without considering the (almost) axial vector-like interaction of $Z' c\bar{c}$, the constraint (\ref{eq:jpex}) provides 
\dis{
|g_{sb}^R g_{cc}^{L,R} \sin\theta_R| &< \mathcal{O}(10^{-5})~,\\
|g_{sb}^L g_{cc}^{L,R} \sin\theta_L| &< 10^{-4}~.
\label{eq:phiscon}
}
When $\theta_L = \theta_R$ or one of the couplings $|g_{sb}^L|$ and $|g_{sb}^R|$ is dominant, the angle $|\sin(2\tilde{\sigma}_s)| \approx |\sin\theta_L|$ or $|\sin\theta_R|$. In this case, hence, we can directly use the constraint (\ref{eq:jpex}) to check the allowed parameter space for the dimuon charge asymmetry.   

In the mean time, we can also analyze more general case that the simple assumption $h_s \approx 0$ is not applied, while $h_s$ should still satisfy the $\Delta M_s$ bound as Fig. \ref{fig:delms}. Then, the NP contribution in $\phi_s^{J/\psi\,\phi}$ is small when there is a fine cancellation between the first and second terms in (\ref{eq:jppanal}). From (\ref{eq:finetune}), we know that the off-diagonal couplings $g_{sb}^{L,R}$ must be around the asymptotic lines (\ref{eq:asymline}) to satisfy $\Delta M_s$ bound unless both of them are smaller than $10^{-3}$. Since the condition (\ref{eq:asymline}) demands $\theta_L = \theta_R$ which makes the various constraints simpler, we can proceed our analysis according to the values of the off-diagonal couplings. For the clear readability of our paper, we leave the detail explanation in our Appendix \ref{appen:hs}. One thing to stress is that our numerical analysis in the {\it $g_{cc}$ scenario} will be proceeded with the conservative assumption $h_s \approx 0$ but our result can be generally applied even when a fine cancellation between the first and second terms in (\ref{eq:jppanal}) exists.  

\begin{figure}
\includegraphics[width=9cm]{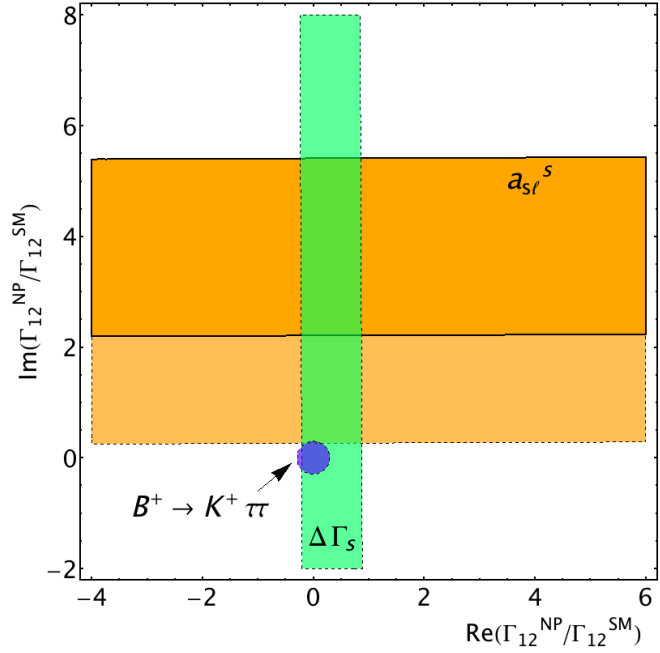}
\caption{This shows the consistency of explaining the observed $a_{s\ell}^s$ within $1\sigma$. The orange color with thick line boundary region denotes $-a_{s\ell}^s/(a_{s\ell}^s)^{\rm SM} > 580$ and the dashed line $68$ when $h_s = |M_{12}^{\rm NP} / M_{12}^{\rm SM}|\ll 1$ such as the {\it $g_{\tau \tau}$ scenario}. The experimental results with 90\%C.L. $\Delta \Gamma_s$ at the LHCb 1fb$^{-1}$ is shown with light green color, while the result of $B^+ \to K^+ \tau^+ \tau^-$  from \cite{Bobeth:2011st} is shown in the light purple color. Actually this bound also covers other bounds such as $B_s \to \tau^+ \tau^-$, $B \to X_s \tau^+ \tau^-$, $B \to X_s \gamma$, $B \to X_s \ell^+ \ell^-$, and $B \to K^{(\ast)} \ell^+ \ell^-$. (Among them, the strongest bound is given by $B^+ \to K^+ \tau^+ \tau^-$ as seen in \cite{Bobeth:2011st}.) 
Figure is simply depicted in terms of Re($\Gamma_{12}^{s\,{\rm NP}}/\Gamma_{12}^{s\,{\rm SM}}) = \tilde{h}_s \cos2\tilde{\sigma}_s$ and Im($\Gamma_{12}^{s\,{\rm NP}}/\Gamma_{12}^{s\,{\rm SM}}) = \tilde{h}_s \sin2\tilde{\sigma}_s$ with the assumption that $h_s \ll 1$. We can easily see that the explanation of $a_{s\ell}^s$ and the experimental bound $\Delta \Gamma_s$ are orthogonal since they depend on the Im.($\Gamma_{12}^{\rm NP}$) and Re.($\Gamma_{12}^{\rm NP}$), respectively. The three experimental results are only marginally consistent.  
}
\label{fig:tau2}
\end{figure}

Now, we move to the issue of $\Delta \Gamma_s$. The analytic expression of $\Delta \Gamma_s / (\Delta \Gamma_s)^{\rm SM}$ is given as
\dis{
\frac{\Delta \Gamma_s}{(\Delta \Gamma_s)^{\rm SM}} &= \frac{2|\Gamma_{12}| \cos\phi_s}{2|\Gamma_{12}^{\rm SM}| \cos\phi_s^{\rm SM}} \\
&= \frac{1}{\sqrt{1+h_s^2+2h_s\cos2\sigma_s}} \left[ (1+h_s \cos 2\sigma_s)(1+\tilde{h}_s\cos2\tilde{\sigma}_s) + h_s \tilde{h}_s \sin2\sigma_s \sin2\tilde{\sigma}_s \right.\\
&\hspace{2cm} \left. - \tan\phi_s^{\rm SM} \left( h_s \sin2\sigma_s (1+\tilde{h}_s\cos2\tilde{\sigma}_s) - \tilde{h}_s \sin 2\tilde{\sigma}_s (1+h_s \cos2\sigma_s) \right) \right]~.
\label{eq:delgamanal}
}
With Eq. (\ref{eq:ra}), we can see that the enhancement of dimuon charge asymmetry is always possible without suffering from the constraint on $\Delta \Gamma_s / (\Delta \Gamma_s)^{\rm SM}$. This is because the enhancement of $a_{q\ell}$ is from Im($\Gamma_{12}$) and that of $\Delta \Gamma_q$ from Re($\Gamma_{12}$) along the direction of Re($M_{12}$), as easily expected from the first relation in (\ref{eq:asanal}). The consistent parameter space is shown with 2D plot as our Fig. \ref{fig:tau2}, where the parameter space is free from the $\Delta M_s$ bound.

On the other hand, the other constraints from \cite{Bobeth:2011st} such as $B^+ \to K^+ \tau^+ \tau^-$, $B_s \to \tau^+ \tau^-$, $B \to X_s \tau^+ \tau^-$, $B \to X_s \gamma$, $B \to X_s \ell^+ \ell^-$, and $B \to K^{(\ast)} \ell^+ \ell^-$ provide additional interesting limit in the allowed parameter space. (Among them, the strongest bound is given by $B^+ \to K^+ \tau^+ \tau^-$.) The experimental bounds can be analytically expressed with $\tilde{h}_s \cos2\tilde{\sigma}_s$ and $\tilde{h}_s \sin2\tilde{\sigma}_s$, in addition to the dimuon charge asymmetry value in (\ref{eq:ratio}). The bound is $|\Gamma_{12}^{s\,{\rm NP}}/\Gamma_{12}^{s\,{\rm SM}}| < 0.3$ from \cite{Bobeth:2011st}. Therefore, it is possible to check the consistency of $a_{s\ell}^s$, $\Delta \Gamma_s$, and $B^+ \to K^+ \tau^+ \tau^-$ in terms of such parameters as Fig. \ref{fig:tau2}. Consequently, the three experimental results are only marginally consistent at the region allowing large NP contribution in $a_{s\ell}^d$.

\subsection{$b \to s \nu \bar{\nu}$}
\label{sec:nu}

\begin{figure}
\includegraphics[width=9cm]{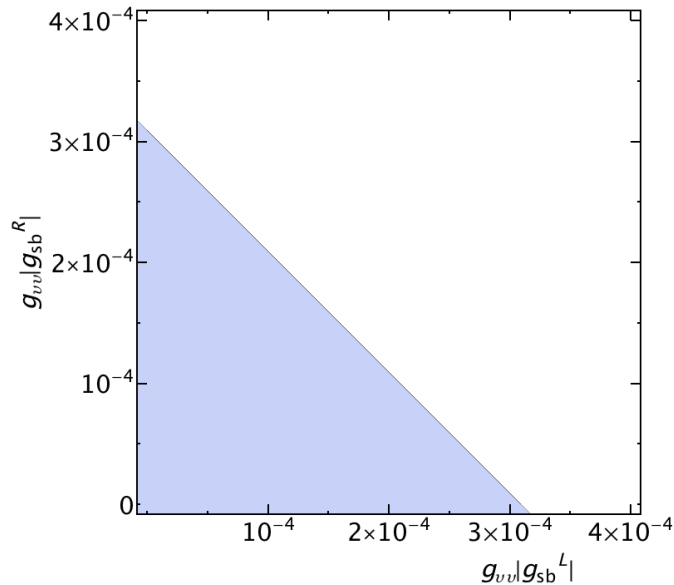}
\caption{The colored region denotes the parameter space allowed by the 90\% C.L. experimental bounds of $g_{\nu \nu} |g_{sb}^{L,R}|$. This is the case that $\theta_L = \theta_R \equiv \theta = \pi/4$ and $g_{\nu \nu}>0$. This parameter space is free from the $\Delta \Gamma_s$ bound by making $\Gamma_{12}^{s\,{\rm NP}}/\Gamma_{12}^{s\,{\rm SM}}$ almost imaginary, as well as the $\Delta M_s$ bound. Since $g_{\tau \tau}^L = g_{\nu \nu}$, the rough upper limit of the coupling is obtained $g_{\tau \tau}^L|g_{sb}^{L,R}| \lesssim 3 \times 10^{-4}$. Even for the other cases, the upper limit is below $10^{-3}$.}
\label{fig:nu}
\end{figure}

In the case that the non-zero $g_{sb}^{L,R} g_{\tau \tau}^{L}$ provides the sizable enhancement in $a_{s\ell}^s$, the coupling $g_{\tau \tau}^L$ is constrained by its partner in the SU(2) doublet $g_{\nu \nu}^L \equiv g_{\nu \nu}$.
Following the analysis in \cite{arXiv:1111.1257,Altmannshofer:2009ma}, we can obtain the limit of $g_{\nu \nu} g_{sb}^{L,R}$ from $B \to K^{\ast} \nu \bar{\nu}$, $B \to K \nu \bar{\nu}$, and $B \to X_s \nu \bar{\nu}$. The detail way of calculating the $Z'$ contribution in these processes are well summarized in our Appendix \ref{appen:nu}.

As experimental upper bounds at 90\% C.L. ($1.65 \sigma$), we obtain from \cite{PDG2011,hep-ex/0010022} such that
\begin{eqnarray}
{\rm Br}(B \to K^{\ast} \nu \bar{\nu}) &<& 8 \times 10^{-5} \hspace{0.5cm} \text{\cite{PDG2011}}~, \\
{\rm Br}(B \to K \nu \bar{\nu}) &<& 1.3 \times 10^{-5} \hspace{0.5cm} \text{\cite{PDG2011}}~, \\
{\rm Br}(B \to X_s \nu \bar{\nu}) &<& 6.4 \times 10^{-4} \hspace{0.5cm}  \text{\cite{hep-ex/0010022}}~.
\end{eqnarray}
Combining all the limits, we obtain the limit of the couplings as Fig. \ref{fig:nu}. The allowed range of $g_{\nu \nu} |g_{sb}^{L,R}|$ from the 90\% C.L. experimental bounds is shown. We deal with the case that $\theta_L = \theta_R \equiv \theta$, which is considered in the fine-tuned region of (\ref{eq:asymline}). This figure is an example $\theta=\pi/4$ and $g_{\nu \nu}>0$. The rough upper limit of the coupling is obtained $g_{\nu \nu}|g_{sb}^{L,R}| <  3\times10^{-4}$. Even for the other cases, the upper limit is below $10^{-3}$.

\begin{figure}
\subfigure[\ Possible enhancement in $a_{s\ell}^s$]{
\includegraphics[width=7.5cm]{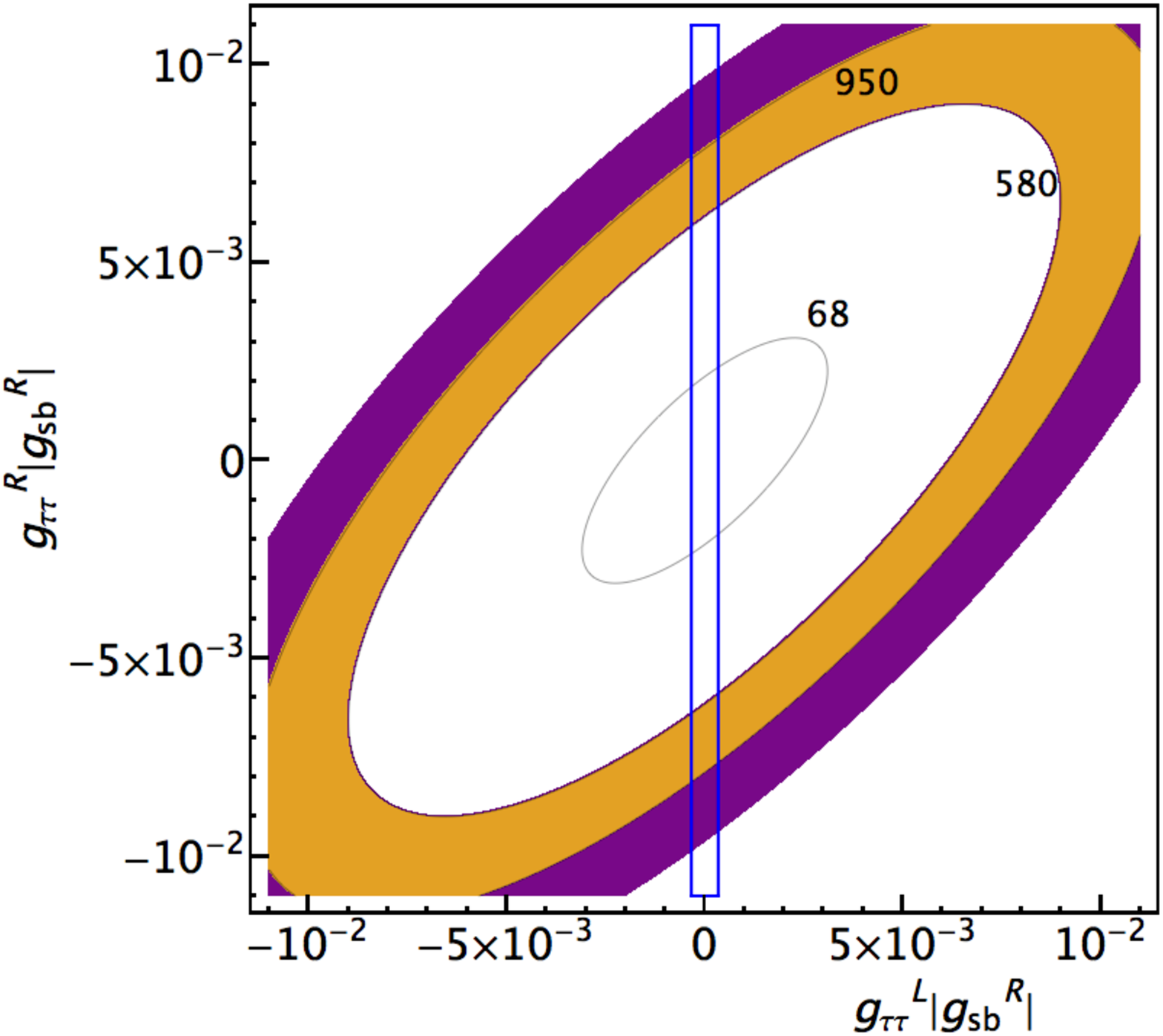}}
\quad
\subfigure[\ Limit of $g_{\tau \tau}^L/g_{\tau \tau}^R$]{
\includegraphics[width=7.5cm]{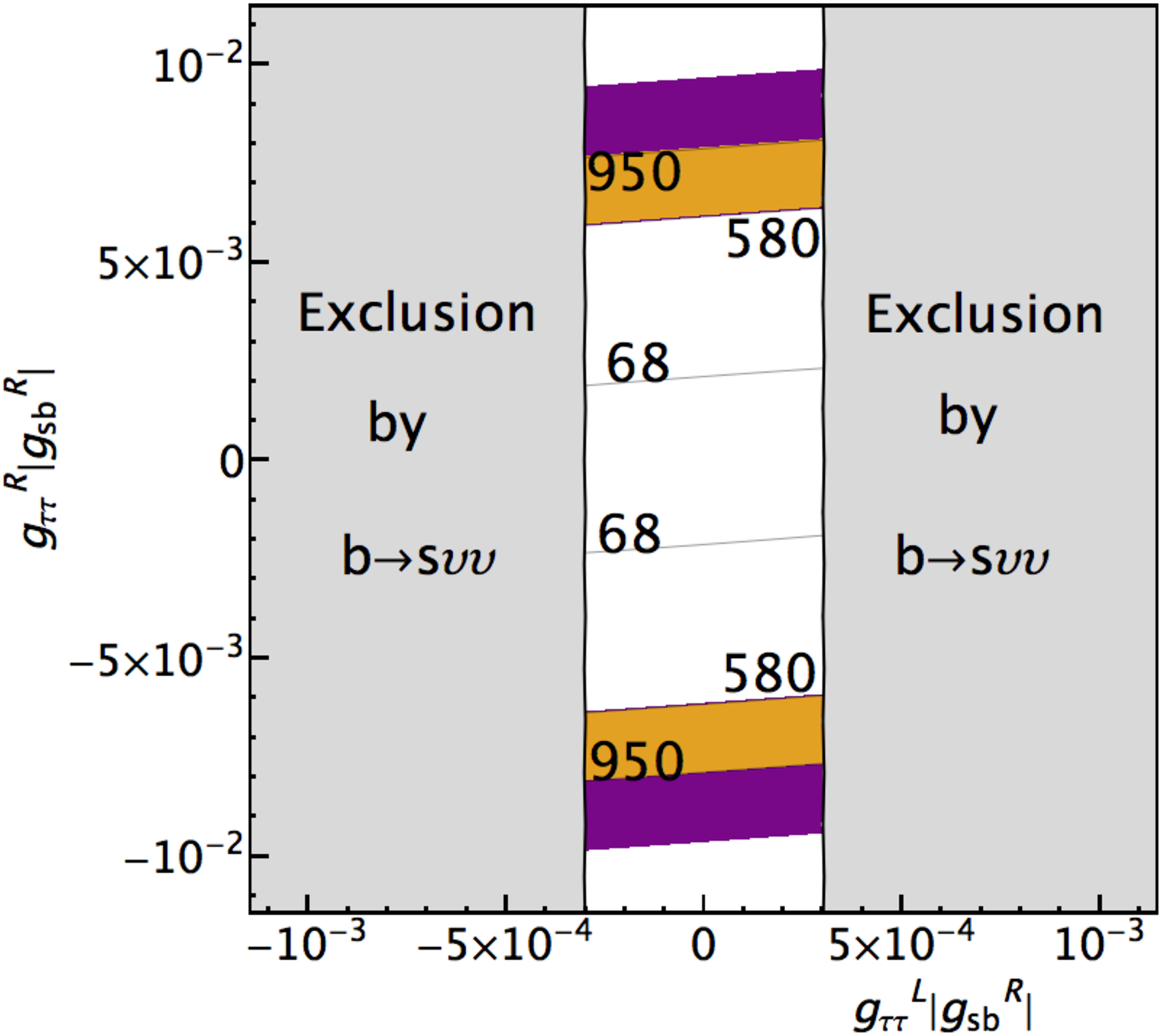}
}
\caption{The limit of the couplings explaining the observed $a_{s\ell}^s$ in the {\it $g_{\tau \tau}$ scenario} is shown. In (a), the region inside the blue line box is those remained after applying the $b\to s \nu \bar{\nu}$ constraint, which is precisely shown in (b). In figure (b), we expressed the conservative exclusion region (grey region) based on the experimental bounds of 90\% C.L from the $b \to s \nu \bar{\nu}$ processes, which is $|g_{\tau \tau}^{L} g_{sb}^{L,R}| < 3 \times 10^{-4}$. This is the fine-tuned case that $g_{sb}^R = a g_{sb}^L$ and $\theta_L = \theta_R = \pi/4$ as an example. This parameter space is free from the $\Delta \Gamma_s$ bound by making $\Gamma_{12}^{s\,{\rm NP}}/\Gamma_{12}^{s\,{\rm SM}}$ almost imaginary, as well as the $\Delta M_s$ bound. The ratio of $-a_{s\ell}^s/(a_{s\ell}^s)^{\rm}$ is shown as a contour plot with the contours 68, 580, and 950. To explain the asymmetry within $1\sigma$, the value of $|g_{\tau \tau}^R|$ must be much larger than $|g_{\tau \tau}^L|$.}
\label{fig:tau1}
\end{figure}

In the $g_{\tau \tau}^L$ scenario, this provides a strong direct upper bound of the couplings as shown in Fig. \ref{fig:tau1}. To explain the asymmetry within $1\sigma$, the value of $|g_{\tau \tau}^R|$ must be much larger than that of $|g_{\tau \tau}^L|$. {\footnote{In this case, the anomaly cancellation in the {\it $g_{\tau \tau}$ scenario} is threaten, unless we assume a scenario like the effective $Z'$ model \cite{Fox:2011qd}. This is because there is no way to cancel the SU(2)$^2$U(1)$^\prime$ anomaly from the $g_{\tau \tau}^R$ coupling.}}

\subsection{$\sin2\beta$ from $B^0 \to J/\psi K_S$}
\label{sec:sin2b}

In this section, we deal with the additional experimental bound when the NP phases contribute to the $b \to scc$ process, such as the {\it $g_{cc}$ scenario}. This is the indirect CP asymmetry $\sin2\beta$ in the ``golden plate" mode $B \to J/\psi K_S$. The SM prediction of $\sin2\beta$ can be  obtained from the fit of the unitarity triangle. According to \cite{Charles:2004jd}, we obtain $\sin(2\beta)^{\rm fit} = 0.731 \pm 0.038$, while the experimental measurements provide $\sin2\beta^{\rm meas} = 0.668 \pm 0.028$. In this case, the SM prediction is within $1\sigma$ of the measured value. The detail analytic form of $\sin2\beta$ is well described in \cite{Chiang:2009ev} and \cite{Alok:2010ij}, which is similar to $\sin2\beta_s$ in $B_s \to J/\psi\,\phi$ as (\ref{eq:phisjp}). In the absence of the SM strong phase, 
\dis{
\sin2\beta^{\rm meas} = \sin(2\beta)^{\rm fit} + 2|r| \cos (2\beta)^{\rm fit} \sin\varphi ~.
}
As (\ref{eq:phisjp}), this relation is obtained when $|r| \ll 1$ and we use the exact relation in \cite{Chiang:2009ev} in our figures.

In the {\it $g_{cc}$ scenario}, the analytic form of $|r|$ is obtained as 
\begin{eqnarray}
|r| &=& \left| \frac{1}{g_1^2} \frac{M_Z^2}{M_{Z'}^2} \frac{2(g_{cc}^L + g_{cc}^R)(g_{sb}^L + g_{sb}^R)}{V_{cb} V_{cs}^{\ast} \cdot 0.17} \right| 
\approx (5.2 \times 10^2) \times |(g_{cc}^L + g_{cc}^R)(g_{sb}^L + g_{sb}^R)|~, 
\end{eqnarray}
and the angle $\varphi$ is simply obtained in the fine-tuned case (\ref{eq:asymline}) such that $\varphi = \theta, \theta+\pi$. Therefore, the allowed range with 90\% C.L. of the experimental result and the SM fit is obtained as
\dis{
-1.4 \times 10^{-4} < |(g_{cc}^L + g_{cc}^R)(g_{sb}^L + g_{sb}^R)| \sin\varphi < 1.4 \times 10^{-5}~.
\label{eq:sin2b1}
}
As the experimental bound by $\phi_s^{J/\psi\,\phi}$, this provides the strong constraint on the NP parameter space unless the coupling $Z'c\bar{c}$ is (almost) axial vector-like. This bound will be shown in Sec. \ref{sec:gcc} with other experimental constraints.

On the other hand, the fitting value of $\sin(2\beta)^{\rm fit}$ is enlarged if we drop the value of $|V_{ub}|$ as an input since its inclusive and exclusive determination has a large difference. Instead, it is possible to use as inputs from the experiments, $\epsilon_K$, $\Delta M_s / \Delta M_d$, Br.($B \to \tau \nu$). In this case, we obtain $\sin(2\beta)^{\rm fit} = 0.867 \pm 0.048$ which induces more than $3\sigma$ deviation from the observed central value \cite{demiseCKM}. By doing this, we can accommodate sizable NP contribution to $\sin2\beta$ by $g_{sb}^{L,R} g_{cc}^{L,R}$ without sizable deviations in the $B \to \tau \nu$ branching ratio and $\epsilon_K$ \footnote{In contrast to this interesting approach, it is fair to note that the Belle collaboration recently updated their result on Br($B^- \to \tau^- \bar{\nu}_\tau$) which is consistent with the usual global fit to the Cabbibo-Kobayashi-Maskawa matrix elements \cite{Adachi:2012mm}.  
}
. In this case, the value of $|r|$ from the NP contribution is allowed up to $20.0+6.5 = 26.5$ \% with the $1\sigma$ predictions. In terms of the {\it $g_{cc}$ scenario}, the allowed range with 90\% C.L. of the experimental result and the SM fit in this case induces
\dis{
-2.8 \times 10^{-4} < |(g_{cc}^L + g_{cc}^R)(g_{sb}^L + g_{sb}^R)| \sin\varphi < -1.0 \times 10^{-4}~. 
\label{eq:sin2b2}
}
The corresponding parameter region will be discussed in Sec. \ref{sec:gcc}.

\section{$g_{\tau \tau}$ {\it scenario} for the dimuon charge asymmetry}
\label{sec:gtata}

In this section, we explore the possible parameter space of the {\it $g_{\tau \tau}$  scenario} to explain the like-sign dimuon charge asymmetry, combined with the experimental bounds discussed in the previous section. In this scenario, the enhancement of $\Gamma_{12}^2$ is realized in the process of the $\tau$ loop-induced $Z'$ exchange. As seen in Fig. \ref{fig:tauhs}, we can simply assume $h_s \ll 1$ for the rough analysis. Then, the ratio of the flavor specific asymmetry 
\dis{
a_{s\ell}^s / (a_{s\ell}^s)^{\rm SM} &=   - \tilde{h}_s \sin2\tilde{\sigma}_s \cot\phi_s^{\rm SM} + 1 + \tilde{h}_s \cos2\tilde{\sigma}_s  \\
&\approx - (2.6 \times 10^2)~ \tilde{h}_s \sin2\tilde{\sigma}_s + 1 + \tilde{h}_s \cos2\tilde{\sigma}_s~,
\label{eq:ratio}
}
where we put the central value of $\phi_s^{\text{SM}} = 3.8 \times 10^{-3}$. Due to the strong LHCb constraint on $\Delta \Gamma_s$, the term $- (2.6 \times 10^2)~ \tilde{h}_s \sin2\tilde{\sigma}$ is dominant so that $\sin2\tilde{\sigma}_s$ is far from 0. The value of $\tilde{h}_s$ in the $g_{\tau \tau}$ scenario is obtained from \cite{Alok:2010ij} such that
\dis{
\tilde{h}_s \approx (6.7 \times 10^3) \times {\rm Abs.}&\left[((g_{sb}^L)^2 + (g_{sb}^R)^2)\left\{1.1 g_{\tau \tau}^L g_{\tau \tau}^R - 0.5((g_{\tau \tau}^L)^2 + (g_{\tau \tau}^R)^2)\right\}  \right. \\
& \left. + g_{sb}^L g_{sb}^R \left\{-3.3 g_{\tau \tau}^L g_{\tau \tau}^R +1.0((g_{\tau \tau}^L)^2 + (g_{\tau \tau}^R)^2) \right\}\right]~.
\label{eq:thstau}
}
 
\begin{figure}
\includegraphics[width=9cm]{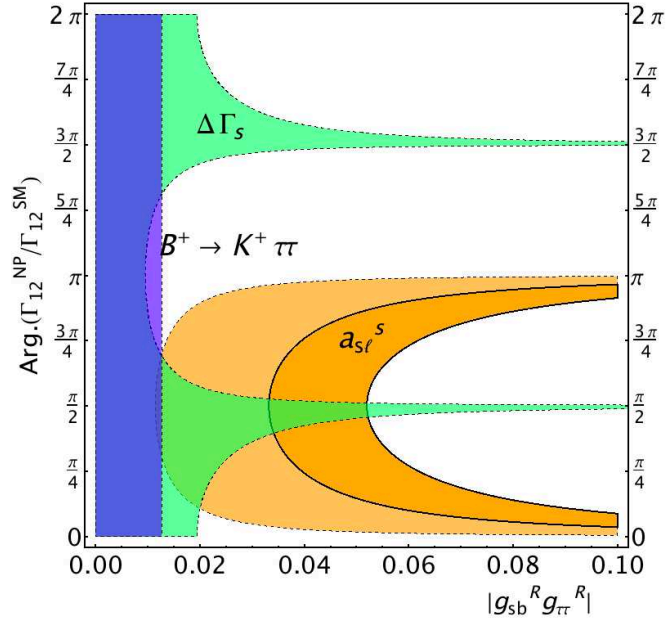}
\caption{We changed the parameters in Fig. \ref{fig:tau2} in terms of $2\tilde{\sigma}_s - |g_{sb}^{R} g_{\tau \tau}^R|$ in the conservative case that $g_{\tau \tau}^L = 0.1 g_{\tau \tau}^R$ and $g_{sb}^R =a g_{sb}^L$ ($h_s \approx 0$). The description on the colored region is same as that in Fig. \ref{fig:tau2}. This shows better understanding on the limits of the couplings in the {\it $g_{\tau \tau}$ scenario}. We see that the rough consistent region of $|g_{sb}^R g_{\tau \tau}^R|$ is about $10^{-2}$ with $a_{s\ell}^d / (a_{s\ell}^d)^{\rm SM} = 21$.
}
\label{fig:tautest}
\end{figure}

To obtain the enhancement in $a_{s\ell}^s$ as large as our second reference point $(1,-580)$ in (\ref{eq:base}), the rough lower limit of the couplings inside the Abs. symbol must be $3.3 \times 10^{-4} = (1.8 \times 10^{-2})^2$ to for $\sin2\tilde{\sigma}_s =1$. To obtain the enhancement as our first reference point $(a_{s\ell}^d/(a_{s\ell}^d)^{\rm SM}, a_{s\ell}^s/(a_{s\ell})^{\rm SM}) = (21,-68)$, the limit lowers to $3.9 \times 10^{-5} = (6.2 \times 10^{-3})^2$. Consequently, we roughly obtain the limit of the dominant new coupling to explain the dimuon charge asymmetry within $1\sigma$ in the $\chi^2$-fit 
\dis{
|g_{sb}^{L,R} g_{\tau \tau}^{L,R}| > 1.8 \times 10^{-2}~&{\rm without}~(a_{s\ell}^d)^{\rm NP}~,\\
|g_{sb}^{L,R} g_{\tau \tau}^{L,R}| > 6.2 \times 10^{-3}~&{\rm with}~a_{s\ell}^d/(a_{s\ell}^d)^{\rm SM} = 21~.
\label{eq:gccresult}
} 
Since $|g_{\tau \tau}^{L,R}| < 1$, the values of Max\{$|g_{sb}^{L,R}|$\} cannot be smaller than $\sim 6.2 \times 10^{-3}$ so that the couplings $g_{sb}^{L,R}$ lie on the asymptotic lines (\ref{eq:asymline}), having more than 1\% fine tuning. The allowed parameter space is shown in Fig. \ref{fig:tautest} where we used $g_{\tau \tau}^L =0.1 g_{\tau \tau}^R$ to maximally satisfy the constraint by $b \to s\nu \bar{\nu}$ as explained in Sec. \ref{sec:nu}. We see that the rough consistent region of $|g_{sb}^R g_{\tau \tau}^R|$ is about $10^{-2}$ with $a_{s\ell}^d / (a_{s\ell}^d)^{\rm SM} = 21$.  We also show the allowed parameter space of $\Delta \Gamma_s$ and $B^+ \to K^+ \tau^+ \tau^-$ from \cite{Bobeth:2011st}, redrawn from the allowed region in Fig. \ref{fig:tau2}.

Consequently, the {\it $g_{\tau \tau}$ scenario} where the $Z'$ coupling to the $\tau$ pair enhances the $a_{s\ell}^s$ requires the existence of the coupling $|g_{sb}^{L,R} g_{\tau \tau}^{L,R}|$ larger than about $10^{-2}$ to explain the dimuon charge asymmetry. Therefore, this parameter space cannot avoid the fine tuning from the $\Delta M_s$ constraint. In addition, due to the constraint from the $b \to s \nu \bar{\nu}$ experiments, the coupling $|g_{\tau \tau}^L|$ must be as small as $3 \times 10^{-4}$. This result demands a non-trivial approach in establishing an anomaly free model as mentioned at the end of Sec. \ref{sec:nu}. The allowed parameter space explaining the observed $a_{s\ell}^s$ within $1\sigma$ is marginally consistent with the experimental bounds of $\Delta \Gamma_s$ at the LHCb 1fb$^{-1}$ and the result of $B^+ \to K^+ \tau^+ \tau^-$ from \cite{Bobeth:2011st}.

\section{$g_{cc}$ {\it scenario}}
\label{sec:gcc}

\begin{figure}
\includegraphics[width=9cm]{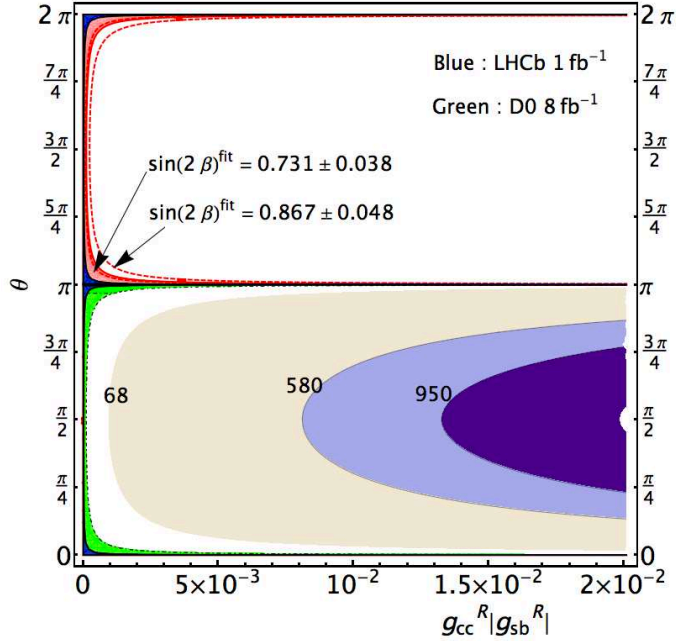}
\caption{We represent the allowed parameter space explaining the dimuon charge asymmetry by fixing $g_{cc}^L=0$ and $g_{cc}^R >0$ on the fine tuning region $g_{sb}^R = a g_{sb}^L$. (Therefore, $\theta_L = \theta_R = \theta$.)  
The parameter region $g_{cc}^R |g_{sb}^R| > 10^{-2}$ is not considered to avoid the rough constraint from $\bar{B}^0 \to D^+ D_s^-$ \cite{Zupanc:2007pu}. Even though we allow the NP contribution which is about half of the SM tree level prediction, it is roughly $g_{cc}^R |g_{sb}^R|/g_1^2 < 0.5 |V_{cb} V_{cs}^{\ast}| \sim 0.02$. 
The numbers in the contours denote the ratio $-a_{s\ell}^s/(a_{s\ell}^s)^{\rm SM}$ as previous figures. The light pink region denotes the 90\% bound from the $B^0 \to J/\psi K_S$ considering the usual fit $\sin(2\beta)^{\rm fit}=0.731 \pm 0.038$ and the area surrounded by the red dashed line is for the special fit $\sin(2\beta)^{\rm fit} = 0.867 \pm 0.048$ in \cite{demiseCKM}. The blue region is 90\% of $\phi_s^{J/\psi\,\phi}$ at the recent 1.0 fb$^{-1}$ LHCb. The minimum value of $|g_{cc}^R g_{sb}^R|$ to explain the D0 dimuon charge asymmetry within $1\sigma$ is about $8 \times 10^{-3}$ and $10^{-3}$, without the NP contribution to $a_{s\ell}^d$ and with the maximal contribution of $a_{s\ell}^d/(a_{s\ell}^d)^{\rm SM} = 21$, respectively. These bounds do not satisfy the experimental constraints, which is expected in our simple analysis. 
}
\label{fig:fcc2}
\end{figure}

In this section, we explore the possible parameter space of the {\it $g_{cc}$  scenario} to explain the like-sign dimuon charge asymmetry, combined with the experimental bounds discussed in the Sec. \ref{sec:ex}. The enhancement of $\Gamma_{12}^s$ from the interference of the SM process and the $Z'$ induced tree level FCNC in $b \to s c\bar{c}$ is calculated from \cite{Golowich:2006gq,Chen:2007dg} such that
\dis{
\Gamma_{12}^{\rm SM+Z'} &= - \frac{m_b^2}{3\pi(2 m_{B_s})}  G^2_F V_{cb} V^{\ast}_{cs} \frac{1}{g_1^2} \frac{M_Z^2}{M_{Z'}^2} K_1 \sqrt{1-4x_c} \\
&\hspace{3cm} \times \left[ 4 g_{sb}^L g_{cc}^L \left\{ (1-x_c) \langle \mathcal{O}_{LL} \rangle + (1+2x_c) \langle \tilde{\mathcal{O}}_{RR} \rangle \right\} \right. \\
&\hspace{3.5cm} + 4 g_{sb}^R g_{cc}^L \left\{ (1-x_c) \langle \mathcal{O}_{LR} \rangle  + (1+2x_c)  \langle \tilde{\mathcal{O}}_{RL} \rangle \right\}\\
&\hspace{3.5cm} \left. + 12x_c g_{sb}^L g_{cc}^R  \langle \mathcal{O}_{LL} \rangle + 12x_c g_{sb}^R g_{cc}^R \langle \mathcal{O}_{LR} \rangle \right]~,
}
where $x_c \equiv m_c^2 / m_b^2$ and $K_1 = 3.11$ is calculated from the Wilson coefficient of the corresponding operators as in \cite{Golowich:2006gq,Chen:2007dg}. This result is different from that \cite{Alok:2010ij} especially adding the suppression factor $x_c$ at the coefficient of the contribution by $g_{sb}^R g_{cc}^R$.  

\begin{figure}
\includegraphics[width=9cm]{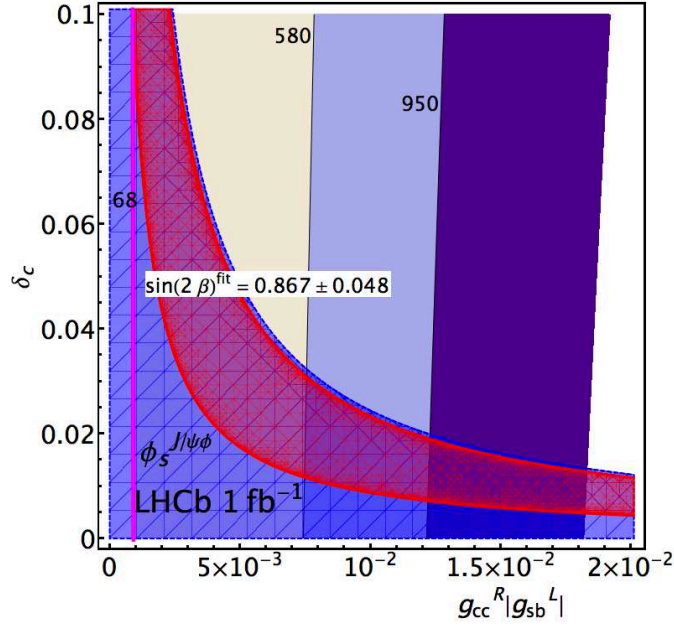}
\caption{We show what extent the interaction $Z'c\bar{c}$ should be axial vector-like in this figure. For various values of the difference $\delta_c \equiv (g_{cc}^L+g_{cc}^R)/g_{cc}^R$, our parameter space explaining the dimoun charge asymmetry is shown for $\delta_c >0$, by fixing $g_{sb}^R \approx (1/a) g_{sb}^L$ with $\theta_L = \theta_R = 3\pi/2$, which is different from the case in Fig. \ref{fig:fcc2}.  
The numbers in the contours denote the value of the ratio $-a_{s\ell}^s/(a_{s\ell}^s)^{\rm SM}$. The contour line with pink color is the boundary $-a_{s\ell}^s/(a_{s\ell}^s)^{\rm SM} = 68$ which demand $a_{s\ell}^d/(a_{s\ell}^d)^{\rm SM} = 21$ to explain the dimoun charge asymmetry within $1\sigma$ in the fit of  Fig. \ref{fig:d0}. The meshed blue region denotes the 90\% allowed region by the recent LHCb 1fb$^{-1}$ result of $\phi_s^{J/\psi\,\phi}$ and the area surrounded by the red dash line is that explaining the $\sin2\beta$ with the special fit $\sin(2\beta)^{\rm fit} = 0.867 \pm 0.048$ proposed in \cite{demiseCKM}. Of course, all the blue region of $\phi_s^{J/\psi\,\phi}$ is also allowed by that using the usual fit $\sin(2\beta)^{\rm fit}=0.731 \pm 0.038$. In this case, the coupling $|g_{cc}^R g_{sb}^R| > 7.5 \times 10^{-3}$ and $10^{-3}$ are required for the explanation of the asymmetry within $1\sigma$ without the NP contribution to $a_{s\ell}^d$ and with the maximal contribution of $a_{s\ell}^d/(a_{s\ell}^d)^{\rm SM} = 21$, respectively. Our parameter space with $-a_{s\ell}^s/(a_{s\ell}^s)^{\rm SM} > 580$ and $a_{s\ell}^d/(a_{s\ell}^d)^{\rm SM}=1$ can simultaneously explain the recent LHCb 1fb$^{-1}$ result and the $\sin2\beta$ for $\delta_c < 2.5 \times 10^{-2}$. 
}
\label{fig:fcc3}
\end{figure}

The value of $\tilde{h}_s$ in the $g_{cc}$ scenario is obtained such that
\dis{
\tilde{h}_s &\approx 173.7 \times \left| \left( 1.15 g_{sb}^L - 1.76 g_{sb}^R \right) g_{cc}^R  + \left( -1.03 g_{sb}^L + 0.64 g_{sb}^R \right) g_{cc}^L\right|~.
\label{eq:thsgcc}
}

When $h_s \ll 1$, the ratio of the flavor specific asymmetry is 
\dis{
a_{s\ell}^s / (a_{s\ell}^s)^{\rm SM} &\approx    -4.6 \times 10^4 \times \left| \left( 1.15 g_{sb}^L - 1.76 g_{sb}^R \right) g_{cc}^R  + \left( -1.03 g_{sb}^L + 0.64 g_{sb}^R \right) g_{cc}^L\right|\sin 2\tilde{\sigma}_s~,
\label{eq:gccra}
}
in the region that the contribution from Re($\Gamma_{12}^{s\,{\rm NP}}/\Gamma_{12}^{s\,{\rm SM}}$) is suppressed to avoid the $\Delta \Gamma_s$ bound. Then, we directly obtain the rough lower limit of the couplings from (\ref{eq:gccra}). To obtain the dimuon charge asymmetry within $1\sigma$ without any NP contribution in $a_{s\ell}^d$, we have
\dis{
\left| \left( 1.15 g_{sb}^L - 1.76 g_{sb}^R \right) g_{cc}^R  + \left( -1.03 g_{sb}^L + 0.64 g_{sb}^R \right) g_{cc}^L\right| \cdot | \sin \theta_{L(R)}| > 1.3 \times 10^{-2}~,
}
while for $(a_{s\ell}^d/(a_{s\ell}^d)^{\rm SM}, a_{s\ell}^s/(a_{s\ell})^{\rm SM}) = (21,-68)$, we have
\dis{
\left| \left( 1.15 g_{sb}^L - 1.76 g_{sb}^R \right) g_{cc}^R  + \left( -1.03 g_{sb}^L + 0.64 g_{sb}^R \right) g_{cc}^L\right| \cdot | \sin \theta_{L(R)}| > 1.4 \times 10^{-3}~.
}
Consequently, we roughly obtain the limit of the dominant new coupling to explain the dimuon charge asymmetry within $1\sigma$ in the $\chi^2$-fit 
\dis{
|g_{sb}^{L,R} g_{cc}^{L,R} \sin \theta_{L(R)}| > 10^{-2}~&{\rm without}~(a_{s\ell}^d)^{\rm NP}~,\\
|g_{sb}^{L,R} g_{cc}^{L,R} \sin \theta_{L(R)}| > 10^{-3}~&{\rm with}~a_{s\ell}^d/(a_{s\ell}^d)^{\rm SM} = 21~.
\label{eq:gccresult}
} 
In this case, the magnitude of $|g_{sb}^{L,R}|$ has less fine tuning from the $\Delta M_s$ constraint compared to that in the {\it $g_{\tau \tau}$ scenario}.  Unless the interaction $Z'c\bar{c}$ is (almost) axial vector-like, the result (\ref{eq:gccresult}) shows a direct contradiction with the constraint by $\phi_s^{J/\psi\,\phi}$ we obtained in (\ref{eq:phiscon}), as well as the $\sin2\beta$ measurements  in Sec. \ref{sec:sin2b}. This result is shown in Fig. \ref{fig:fcc2} in the simple case $g_{cc}^L = 0$ and $g_{sb}^R = a g_{sb}^L$. 

When $|g_{cc}^L + g_{cc}^R| \ll 1$, the constraint by the $\phi_s^{J/\psi\,\phi}$ at the 1.0 fb$^{-1}$ LHCb is loosen. We show what extent the interaction $Z'c\bar{c}$ should be axial vector-like in our Fig. \ref{fig:fcc3}, simultaneously explaining the interesting parameter region of $\sin2\beta$ using $\sin(2\beta)^{\rm fit} = 0.867 \pm 0.048$ \cite{demiseCKM}. As a result, for the explanation of the asymmetry within $1\sigma$ without the NP contribution to $a_{s\ell}^d$, we can find the consistent parameter space $|g_{cc}^R g_{sb}^R| > 7.5 \times 10^{-3}$ and $\delta_c < 2.5 \times 10^{-2}$ when $g_{sb}^R = a g_{sb}^L$ with $\theta_L = \theta_R = 3\pi/2$.

Finally, we discuss the possibility of the model construction providing the (almost) axial vector-like interaction $Z'c\bar{c}$. In this case, we need the sizable coupling $|g_{cc}^L| \sim |g_{cc}^R|$ at least $\mathcal{O}(10^{-2})$. Since the left-handed charm quark constitutes a SU(2) doublet with the left-handed strange quark, the coupling $g_{ss}^L$ is also of $\mathcal{O}(10^{-2})$. Then, we obtain the nonzero off-diagonal couplings $g_{uc}^L$ and $g_{ds}^L$, unless the size of the couplings $g_{uu}^L (g_{dd}^L)$ are same as $g_{cc}^L (g_{ss}^L)$. It must be noted that the size of such off-diagonal couplings are constrained by the bounds such as $K$ or $D$ meson mixings. For example, the $D$ meson mixing provides the strong upper limit of the $g_{uc}^L$ coupling as small as $2 \times 10^{-4}$ \cite{Giudice:2012qq,Altmannshofer:2012ur}. Considering the CKM relation of $g_{uc}^L \approx 0.23 g_{cc}^{L,R}$, the coupling $|g_{cc}^{L,R}|$ must be smaller than $10^{-3}$, which in turn demands $|g_{sb}^{L,R} g_{cc}^{L,R}| < 10^{-3}$. This is out of the $1\sigma$ region for the dimuon charge asymmetry even with arbitrary contribution in $a_{s\ell}^d$, as seen in Fig. \ref{fig:fcc2} and \ref{fig:fcc3}. 

On the other hand, we need to consider the bounds of the $\pi$ production processes from the $B_s$ meson when the size of the couplings $g_{uu}^L = g_{dd}^L$ are almost same as $g_{cc}^L = g_{ss}^L$. For example, the upper bound of Br($B_s \to \pi^+ \pi^-)$ is as strong as $1.2 \times 10^{-6}$ \cite{PDG2011}, while the value of $|g_{uu,dd}^L \ g_{sb}^R|$ is as large as $|g_{cc}^R \ g_{sb}^R| > 7.5 \times 10^{-3}$.
Therefore, the scenario with the (almost) axial vector-like interaction $Z'c\bar{c}$ is not plausible.

Consequently, the {\it $g_{cc}$ scenario} where the $Z'$ coupling to the $c$-quark pair enhances the $a_{s\ell}^s$ requires the existence of the coupling $|g_{cc}^{L,R} g_{sb}^{L,R}|$ larger than $\mathcal{O}(10^{-3})$ to explain the dimuon charge asymmetry. This parameter space has smaller fine tuning from the $\Delta M_s$ compared to the {\it $g_{\tau \tau}$ scenario}, due to the interference with the SM process in the contribution to $\Gamma_{12}^s$. However, the recent LHCb 1fb$^{-1}$ constraint on $\phi_s^{J/\psi\,\phi}$ and $\Delta \Gamma_s$, as well as the constraint from $B \to J/\psi K_S$, is quite strong to demand the (almost) axial vector-like interaction of $Z'c\bar{c}$. On the other hand, the existence of such interaction makes the model construction very hard due to the experimental bounds such as $K$ or $D$ meson mixing and the $\pi$ production from the $B_s$ decays.


\section{Conclusions}
\label{sec:conclusions}

The like-sign dimuon charge asymmetry has been observed at the D0 which is deviated more than $3\sigma$ from the SM prediction. In the recent result in 2011, it was possible to separately detect the flavor specific asymmetry from the $B_s$ and $B_d$ mixing by imposing the impact parameter cut reducing the background. In this paper, we showed that the enhancement of flavor specific asymmetry $a_{s\ell}^s$ is highly constrained by the recent LHCb result with 1fb$^{-1}$ integrated luminosity. We presented the constraints on the $Z'$ couplings $g_{bs}, g_{\tau \tau}$ or $g_{c c}$ and the possible enhancement of $a_{s\ell}^s$. The actual upper bound of the couplings are expressed when $M_{Z'} \approx M_Z$. By simple scaling of the ratio $M_{Z'}/M_Z$, our result can be applied to the other mass of $Z'$ as well.

For the flavor specific asymmetry $a_{s\ell}^s$, there are three kinds of criteria.
By allowing sizable new physics contribution in $B_d$ system ($a_{s\ell}^d$ is 21 times larger than the SM prediction from NP), $|a_{s\ell}^s/a_{s\ell}^{s \ \rm SM}| \ge 68$ is needed to be within $1 \sigma$ region. If there is no new physics in $B_d$ system and the deviation of $A_{s\ell}^b$ is the consequence of $B_s$ system alone, we need $|a_{s\ell}^s/a_{s\ell}^{s \ \rm SM}| \ge 580$ to be within $1\sigma$ region. The central value requires $|a_{s\ell}^s/a_{s\ell}^{s \ \rm SM}| \ge 950$.

The $B_s$ system is highly constrained by the recent LHCb data.
In the absence of the modification in the decay, $\Gamma_{12}^s$, the recent LHCb measurement of $\phi_s^{J/\psi\,\phi}$ strongly constrains the phase of the mixing, $M_{12}^s$. As a result, the maximum enhancement of $a_{s\ell}^s$ from $M_{12}^s$ alone is at most 40 times the SM prediction which is not enough to be within $1\sigma$ even if we allow arbitrary NP contribution to $a_{s\ell}^d$. 

The $b \to c{\bar c} s$ coupling can provide an extra contribution in $\phi_s
^{J\psi\,\phi}$ from the decay.
If the coupling is small enough, the main effect would be to modify the relation between the phase of $M_{12}^s$ and $\phi_s^{J\psi\,\phi}$ and the constraint on $\phi_s^{J\psi\,\phi}$ can be slightly relaxed. In this case the main impact of $b \to c{\bar c} s$ is to avoid the constraints on the phase of $M_{12}^s$ from the LHCb measurement of $\phi_s^{J\psi\,\phi}$.
By allowing the $b \to c{\bar c} s$ coupling, $a_{s\ell}^s$ can be as large as 50 times the SM prediction, which is still smaller than 68 for the $1\sigma$ explanation of the asymmetry with arbitrary $a_{s\ell}^d$.

The $\Gamma_{12}^s$ is constrained by $\Delta \Gamma^s$ measurement which is basically ${\rm Re}( \Gamma_{12}^s)$ when $M_{12}^s$ is almost real.  The enhancement of $a_{s\ell}^s$ is mainly from ${\rm Im} (\Gamma_{12}^s)$ which can affect the other observables like $B^+ \to K^+ \tau \tau$ in the {\it $g_{\tau \tau}$ scenario} and $B_s \to J/\psi\,\phi$ in the {\it $g_{cc}$ scenario}.

The {\it $g_{\tau \tau}$ scenario} where the $Z'$ coupling to the $\tau$ pair enhances the $a_{s\ell}^s$ requires the existence of the coupling $|g_{sb}^{L,R} g_{\tau \tau}^{L,R}|$ larger than about $10^{-2}$ to explain the dimuon charge asymmetry. Therefore, this parameter space cannot avoid the fine tuning from the $\Delta M_s$ constraint $|g_{sb}| \le 10^{-3}$. In addition, due to the constraint from the $b \to s \nu \bar{\nu}$ experiments, $|g_{\tau \tau}^L|$ must be as small as  $3 \times 10^{-4}$. The allowed parameter space explaining the observed $a_{s\ell}^s$ within $1\sigma$ (68 times larger than the SM prediction) is marginally consistent with the experimental bounds of $\Delta \Gamma_s$ at the LHCb 1fb$^{-1}$. 

The {\it $g_{cc}$ scenario} where the $Z'$ coupling to the $c$-quark pair enhances the $a_{s\ell}^s$ requires the existence of the coupling $|g_{cc}^{L,R} g_{sb}^{L,R}|$ larger than about $10^{-3}$ to explain the dimuon charge asymmetry. This parameter space has smaller fine tuning from the $\Delta M_s$ compared to the {\it $g_{\tau \tau}$ scenario}, due to the interference with the SM process in the contribution to $\Gamma_{12}^s$. However, the recent LHCb 1fb$^{-1}$ constraint on $\phi_s^{J/\psi\,\phi}$ and $\Delta \Gamma_s$, as well as the constraint from $B^0 \to J/\psi K_S$, are quite strong. So the interaction $Z'c\bar{c}$ must be (almost) axial vector-like. On the other hand, the existence of $g_{cc}^L$ from the axial vector constraints makes the model construction not plausible due to the experimental bounds such as $K$ or $D$ meson mixing and the $\pi$ production from the $B_s$ decays since $g_{cc}^L = g_{ss}^L$.

Consequently, it is impossible to explain the $1\sigma$ range of like-sign dimuon charge asymmetry using $Z'$ contribution in $B_s$ system without the enhancement in $a_{s\ell}^d$. Even with arbitrary $a_{s\ell}^d$, the {\it $g_{cc}$ scenario} demands unrealistic model construction. Therefore, we need to consider the sizable NP contribution in $a_{s\ell}^d$, while making the $a_{s\ell}^s$ as small as possible. To explain the asymmetry within $1\sigma$ by minimizing the NP contribution to $B_s$ system, we need the $a_{s\ell}^d$ which is only about 21 times the SM prediction at most, as shown in Fig. \ref{fig:d0}. So the required off-diagonal coupling $|g_{db}|$ to enhance the $a_{s\ell}^d$ is smaller than the $|g_{sb}|$. On the other hand, the CKM suppression strengthens the experimental bounds except $\Delta \Gamma_d$ which has been poorly measured so far. The experimental bounds to be analyzed contain $B^0 \to \tau^+ \tau^-$, $B \to \pi \tau^+ \tau^-$, and $B_s^0 \to \bar{K}_0 \tau^+ \tau^-$ when we consider enhancement in $\Gamma_{12}^d$ through nonzero $g_{\tau \tau}$ coupling. For the case with nonzero $g_{cc}$ coupling, we need to consider $B \to J/\psi \pi$ and $B_s \to \bar{K}_0 J/\psi$, etc. Therefore, more careful analysis is required in the future to investigate this approach.

\acknowledgments{We thank Radovan Derm\'i\v sek for the participation of the project from the beginning to the final stage. This work was supported by the NRF of Korea No. 2011-0017051 (HK, SS) and No. 2011-0012630 (SS). SS is also supported by TJ Park POSCO Postdoc fellowship.}

\vskip 0.5cm

\appendix


\section{Calculation of the contribution in to $b \to s \gamma$}
\label{appen:bsg}

\begin{figure}[htbp]
\begin{center}
\includegraphics[width=8cm]{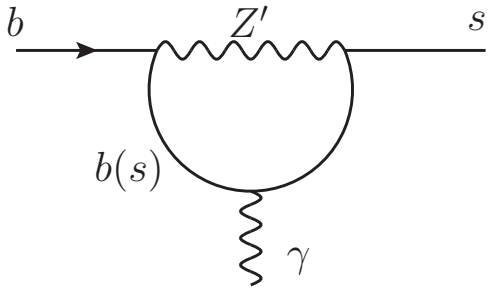}
\end{center}
\caption{$b \to s \gamma$ penguin contribution}\label{fig:bsgam}
\end{figure}

Considering the non-zero value of $g_{bb}^{L,R}$ or $g_{ss}^{L,R}$, we can obtain the upper limit of $|g_{sb}^{L,R}|$ from the $b \to s \gamma$ penguin constraint, shown in Fig. \ref{fig:bsgam}. 

\begin{figure}
\subfigure[\ $\theta_L = \theta_R = \pi/4$]{
\includegraphics[width=7.5cm]{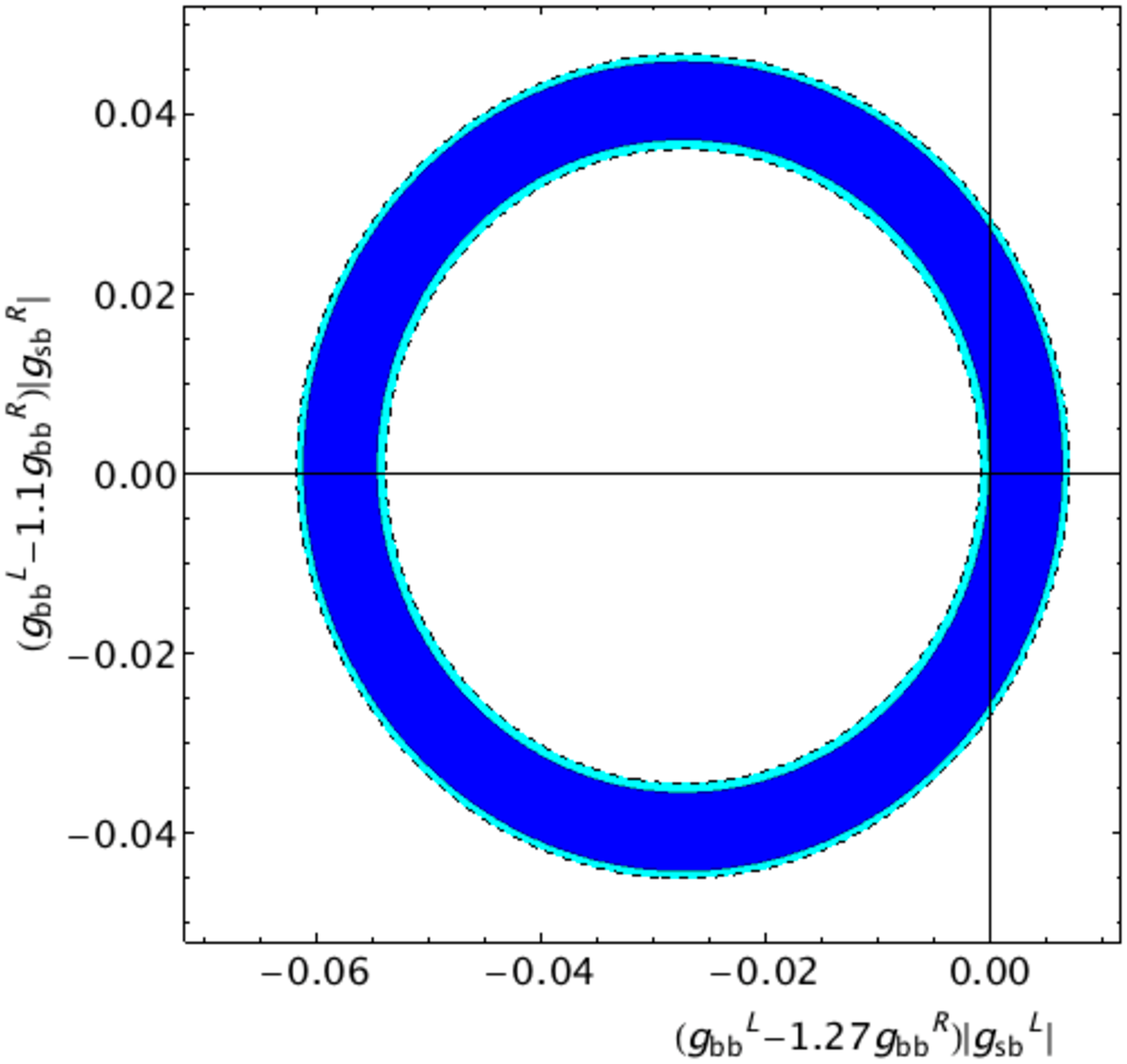}}
\quad
\subfigure[\ $\theta_L = \theta_R = 3\pi/4$]{
\includegraphics[width=7.5cm]{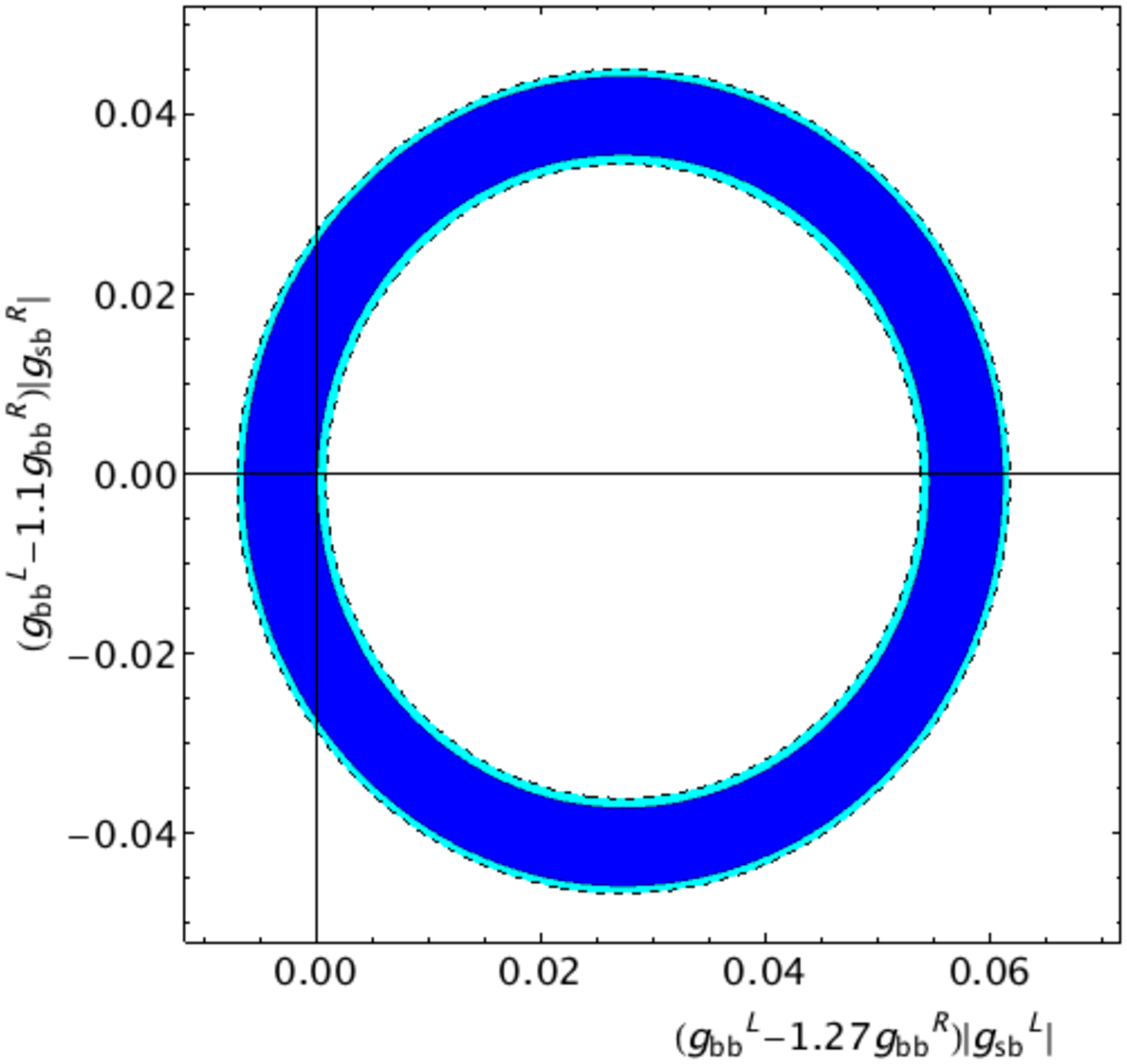}}
\caption{The limit of the couplings from the experimental bounds of 90\% C.L (Blue) and 95\% C.L. (Cyan, Dashed boundary line) of $B \to X_s \gamma$ for fine-tuned cases (a) $\theta_L=\theta_R=\pi/4$ and (b) $\theta_L=\theta_R=3\pi/4$.}
\label{fig:bsg}
\end{figure}

The inclusive decay $\Gamma(B \to X_s \gamma)$ is given approximately by $\Gamma(b \to X_s^{parton} \gamma)$. The nonperturbative correction to this approximation is smaller than the NNLO perturbative QCD corrections to $\Gamma(b \to X_s^{parton} \gamma)$. The theoretical prediction for the partonic $\Gamma(b \to X_s^{parton} \gamma)$ is usually normalized by the semileptonic decay rate to get rid of the uncertainties related to the CKM matrix elements and the fifth power of the $b$-quark mass. Therefore, the SM NNLO result for a photon-energy cut of $E_{\gamma} > 1.6 \gev$ is obtained as \cite{hep-ph/0609232}
\dis{
{\rm Br} (B \to X_s \gamma)_{\rm SM} = (3.15 \pm 0.23) \times 10^{-4} ~,
\label{eq:btosgsm}
}
while the experimental result with for the same energy cut is measured as \cite{arXiv:1010.1589} 
\begin{eqnarray}
{\rm Br} (B \to X_s \gamma)_{\rm exp} = (3.55 \pm 0.24 \pm 0.09) \times 10^{-4} ~.
\label{eq:btosgex}
\end{eqnarray}
The NP contribution in the total branching ratio is below 30\% seeing the result of (\ref{eq:btosgsm}) and (\ref{eq:btosgex}). Therefore, a naive strongest constraint of $|g_{bb}^{L,R} g_{sb}^{L,R}|$ is $< 10^{-2}$ as shown in Fig. \ref{fig:bsg}. However, larger values of the couplings can still satisfy the $b \to s \gamma$ constraint once the coupling ratio $g_{bb}^R / g_{bb}^L$ is about 1.1 or 1.27. The SM NNLO contribution shows a negligible dependence on the value of $\mu_b$. When the LO NP contribution enhances the SM value by 20\%, the $\mu_b$ dependence in the total branching ratio induces about 3\% uncertainty for $\mu_b = 2.5 - 5$ \gev in the numerical analysis in \cite{Buras:2011zb}. 

Following Eq. (5.3) of \cite{Buras:2011zb}, the branching ratio with the NP contribution is obtained
\dis{
{\rm Br} (B \to X_s \gamma) &= (2.47 \times 10^{-3})  \\
& \hspace{0.5cm} \times \left( |C_{7\gamma} ( \mu_b )|^2 + |C'_{7\gamma} ( \mu_b )|^2 + N(E_{\gamma}) \right) ~,
\label{eq:bsgc7}
}
where $N(E_\gamma) = (3.6 \pm 0.6) \times 10^{-3}$ is a nonperturbative contribution. Considering the LO NP contributions 
\dis{
C_{7\gamma} ( \mu_b ) = C_{7\gamma}^{\rm SM} ( \mu_b ) + \Delta C_{7\gamma} ( \mu_b )~,
}
where the central value of the SM contribution is calculated at the NNLO level for $\mu_b = 2.5 \gev$ such that
\dis{
C_{7\gamma}^{\rm SM} ( \mu_b ) = -0.3525~.
}
The NP contribution is obtained as following.
\dis{
\Delta C_{7\gamma} (\mu_b) &= \frac{1}{g_1^2} \frac{M_Z^2}{M_{Z'}^2} \frac{1}{V_{tb} V^{\ast}_{ts}}  \times \left[\left( -\frac29 \kappa_7 + \frac23 \kappa_8 \right) g_{ss}^L g_{sb}^L - 2\kappa_{LL}^s g_{ss}^L (g_{sb}^L)^{\ast}  \right. \\
&\hspace{2cm} + \left(  -\frac29 \kappa_7 + \frac23 \kappa_8 - 2 \kappa_{LL}^b \right) g_{bb}^L (g_{sb}^L)^{\ast}  + \left( \frac23 \kappa_7 - 2 \kappa_8 - 2 \kappa_{LR}^b \right) g_{bb}^R (g_{sb}^L)^{\ast} \\
& \hspace{2.1cm} \left.  + \left( \frac23 \kappa_7 - 2 \kappa_8 \right) \frac{m_s}{m_b} g_{ss}^L g_{sb}^R - 2 \kappa_{LR}^s g_{ss}^R (g_{sb}^L)^{\ast} \right]~,
}
where $\kappa$'s are listed in Table 1 of \cite{Buras:2011zb}. The prime coefficients are obtained as 
\dis{
C_{7 \gamma}^{\prime \rm SM} (\mu_b) = -0.3523~ \frac{m_s}{m_b}~,
}
\dis{
\Delta C'_{7\gamma} (\mu_b) &=  \frac{1}{g_1^2} \frac{M_Z^2}{M_{Z'}^2} \frac{1}{V_{tb} V^{\ast}_{ts}}  \left[\frac{2m_s}{9m_b}\left( -  \kappa_7 + 3\kappa_8 \right)   g_{ss}^R g_{sb}^R - 2\kappa_{LL}^s g_{ss}^R (g_{sb}^R)^{\ast} \right.\\
&\hspace{2cm} + \left(  \frac{2m_s}{9m_b} \left(-\kappa_7 + 3 \kappa_8 \right) - 2 \kappa_{LL}^b \right) g_{bb}^R (g_{sb}^R)^{\ast} \\
&\left. + \left( \frac23 \frac{m_s}{m_b} \kappa_7 - 2 \frac{m_s}{m_b} \kappa_8 - 2 \kappa_{LR}^b \right) g_{bb}^L (g_{sb}^R)^{\ast} + \left( \frac23 \kappa_7 - 2 \kappa_8 \right) \left( \frac{m_s}{m_b} \right)^2 g_{ss}^R g_{sb}^L - 2 \kappa_{LR}^s g_{ss}^L (g_{sb}^R)^{\ast} \right]~.
}
In the simple case that $g_{ss}^{L,R}=0$ and our values of $\kappa$s are not much different from those with the matching scale at around 200 GeV. We obtain the following result,
\begin{eqnarray}
C_{7\gamma} &=& -0.3523 - 9.11 \times ( g_{bb}^L - 1.27 g_{bb}^R )(g_{sb}^L)^{\ast}~, \label{eq:bsgfine1}\\
C'_{7\gamma} &=& -0.3523 \frac{m_s}{m_b} + 6.83 \times (g_{bb}^L - 1.1 g_{bb}^R ) (g_{sb}^R)^{\ast}~. \label{eq:bsgfine2}
\end{eqnarray}

Plugging (\ref{eq:bsgfine1}) into (\ref{eq:bsgc7}) with the consideration of the experimental limit (\ref{eq:btosgex}), we can obtain the limit of $(g_{bb}^L-1.27g_{bb}^R)|g_{sb}^L|$ and $(g_{bb}^L - 1.1 g_{bb}^R) |g_{sb}^R|$ according to a fixed value of the ($\theta_L$, $\theta_R$) such that 
\dis{
&6.42(g_{bb}^L - 1.27 g_{bb}^R) |g_{sb}^L| \cos\theta_L + 82.99 (g_{bb}^L - 1.27 g_{bb}^R)^2 |g_{sb}^L|^2 \\
& \hspace{2cm} -0.11 (g_{bb}^L - 1.1 g_{bb}^R) |g_{sb}^R| \cos\theta_R + 46.65 (g_{bb}^L - 1.1 g_{bb}^R)^2 |g_{sb}^R|^2 = 0.016 \pm 0.01~,
}
within $1 \sigma$ up to $\mathcal{O}(10^{-4})$, calculated with $m_s = 100$ MeV and $m_b = 4.2$ GeV. \\


\section{The effect of the theoretical uncertainties in the form factors}
\label{appen:form}

Considering the square root error propagation, this uncertainty changes the quantity $k_{\pm}$ to 1/3 or twice of the original calculation. (For the Ball-Zwicky model, $k_+  = (8.02 A_1 + 3.35 Z)/(8.02 A_1 - 3.35 Z)$ for the form factors $A_1, Z$ in \cite{Chiang:2009ev}. Considering the 10 \% uncertainties, $k_+ =  (8.02 \cdot 0.42 ( 1 \pm 0.1) + 3.35 \cdot 0.82 ( 1 \pm 0.1 ))/(8.02 \cdot 0.42 ( 1 \pm 0.1) - 3.35 \cdot 0.82 ( 1 \pm 0.1 ))$. Simple calculation with $A_1 = 0.42 (1-0.9)$ and $Z = 0.82 (1+0.1)$ can make $k_+$ very large as 518. However, the error propagation without considering the covariance can make it smaller. 
\dis{
k_+ &= \frac{8.02 \cdot 0.42 + 3.35 \cdot 0.82 \pm \sqrt{(8.02)^2 (0.042)^2 + (3.35)^2 (0.082)^2}}{8.02 \cdot 0.42 - 3.35 \cdot 0.82 \pm \sqrt{(8.02)^2 (0.042)^2 + (3.35)^2 (0.082)^2}} \\
&= \frac{8.02 \cdot 0.42 + 3.35 \cdot 0.82}{8.02 \cdot 0.42 - 3.35 \cdot 0.82}  \left( 1 \pm  \sqrt{(8.02)^2 (0.042)^2 + (3.35)^2 (0.082)^2} \right.\\
&\hspace{2cm} \left. \times \sqrt{\frac{1}{(8.02 \cdot 0.42 + 3.35 \cdot 0.82)^2} + \frac{1}{(8.02 \cdot 0.42 - 3.35 \cdot 0.82)^2}} \right) \\
& = 9.83 (1\pm 0.70)~,
}
which makes $k_+ = 2.9 - 16.7$.) The ratio of each polarized amplitude can be obtained in the CDF data where the transverse amplitude $A_{\parallel,\perp} = (A_+ + A_-)/\sqrt{2}$ \cite{Acosta:2004gt}.


\section{The analysis on $\phi_s^{J/\psi\,\phi}$ when the assumption $h_s \approx 0$ is not applied}
\label{appen:hs}

As commented in the article, we analyze more general case that the simple assumption $h_s \approx 0$ is not applied, while $h_s$ should still satisfy the $\Delta M_s$ bound as Fig. \ref{fig:delms}. Then, the NP contribution in $\phi_s^{J/\psi\,\phi}$ is small when there is a fine cancellation between the first and second terms in (\ref{eq:jppanal}). From (\ref{eq:finetune}), we know that the off-diagonal couplings $g_{sb}^{L,R}$ must be around the asymptotic lines (\ref{eq:asymline}) to satisfy $\Delta M_s$ bound unless both of them are smaller than $10^{-3}$. Since the condition (\ref{eq:asymline}) demands $\theta_L = \theta_R$ which makes the various constraints simpler, we can proceed our analysis according to the values of the off-diagonal couplings. Therefore, we classify the cases as following.

\begin{figure}
\subfigure[\ Case i)]{
\includegraphics[width=7.5cm]{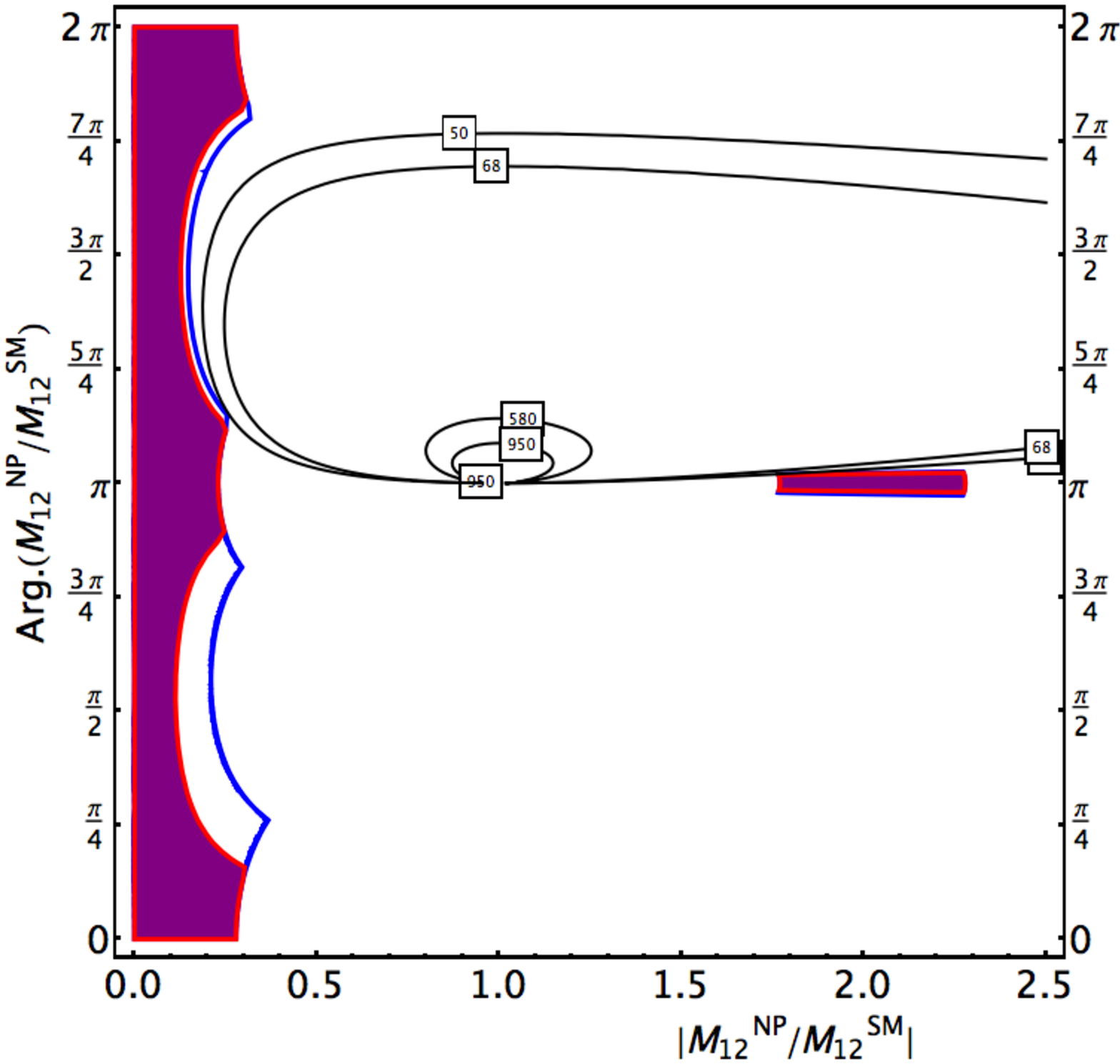}}
\subfigure[\ Case iii)]{
\includegraphics[width=7.8cm]{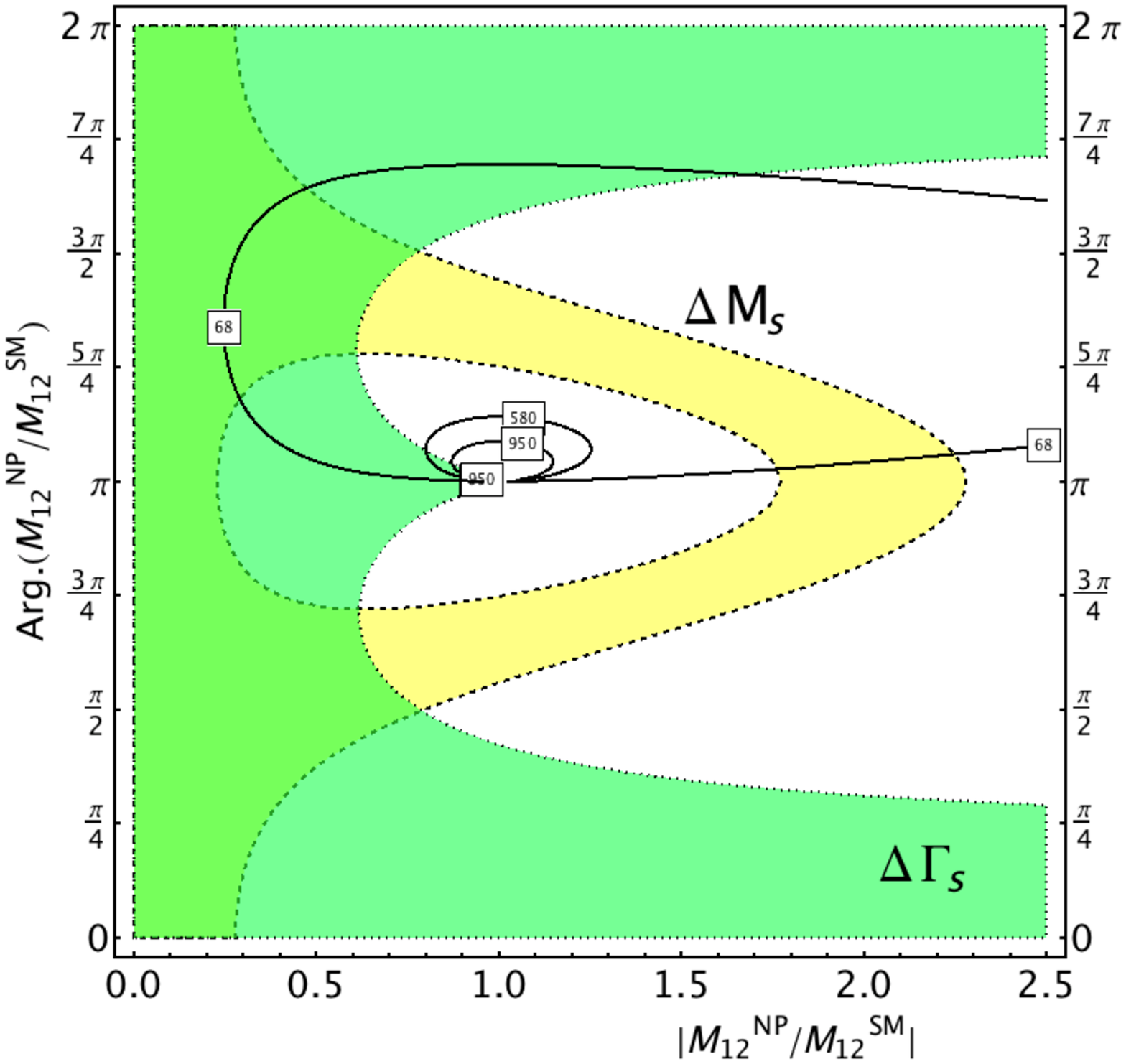}}
\quad
\caption{We represent the allowed parameter space explaining the dimuon charge asymmetry within $1\sigma$ for the case i) (a) and iii) (b) in the contents. (a) The purple region surrounded by the thick red line is the case without the new phase in $b \to sc\bar{c}$ as explained in Fig. \ref{fig:tauhs}. 
When $h_s$ is sizable to cancel the contribution from the $b \to sc\bar{c}$, larger region is allowed by $\phi_s^{J/\psi\,\phi}$ but the NP parameter space is still constrained by $\sin2\beta$ as explained in Sec. \ref{sec:sin2b}. The blue line represents the combined bound of $\Delta M_s$, $\phi_s^{J/\psi\,\phi}$, and $\sin2\beta$ for the case i). 
The contours denote the 50, 68, 580, and 950 of the ratio $-a_{s\ell}^s/(a_{s\ell}^s)^{\rm SM}$. Even with arbitrary contribution in $a_{s\ell}^d$, we see that the combined bound do not allow the enhancement $-a_{s\ell}^s/(a_{s\ell}^s)^{\rm SM} = 68$ to explain the dimuon charge asymmetry within $1\sigma$. (b) The green region surrounded by the dotted line is the 90\% allowed region of $\Delta \Gamma_s$ from the LHCb result (\ref{eq:deltagam}). We can see that the NP contribution in $B_d$ mixing is necessary to explain the dimuon charge asymmetry within $1\sigma$.
}
\label{fig:tauhst2}
\end{figure}

\begin{itemize}
\item[] i) At least one of $|g_{sb}^{L,R}| > 10^{-3}$ but $g_{cc}^{L,R}$ is small enough to ignore $\Gamma_{12}^{\rm NP}/\Gamma_{12}^{\rm SM}$. {\footnote{The coupling $|g_{sb}^{L,R} g_{cc}^{L,R}|$ can be either small or are in a special relation making $\Gamma_{12}$ small.}}
\item[] ii) At least one of $|g_{sb}^{L,R}| > 10^{-3}$ and $g_{cc}^{L,R}$ is large. ($\Gamma_{12}^{\rm NP}/\Gamma_{12}^{\rm SM}$ is sizable.)
\item[] iii) $|g_{sb}^{L,R}| \le 10^{-3}$ and $g_{cc}^{L,R}$ is small enough to ignore $\Gamma_{12}^{\rm NP}/\Gamma_{12}^{\rm SM}$.
\item[] iv) $|g_{sb}^{L,R}| \le 10^{-3}$ but $g_{cc}^{L,R}$ is large enough to make $\Gamma_{12}^{\rm NP}/\Gamma_{12}^{\rm SM}$ sizable .
\end{itemize}

For the case i), the off-diagonal couplings $g_{sb}^{L,R}$ must satisfy the fine tuning condition (\ref{eq:asymline}), which demands $\theta_L = \theta_R$. Then, it is possible to directly apply the constraint of $\sin2\beta$ from $B^0 \to J/\psi K_S$ as explained in Sec. \ref{sec:sin2b}. Considering this effect, it is possible to obtain the limit of the parameters $2\sigma_s$ and $h_s$ as shown in Fig. \ref{fig:tauhst2} (a). In the figure, the combined bound from $\Delta M_s$, $\phi_s^{J/\psi\,\phi}$, and $\sin2\beta$ is represented as the region surrounded by the blue line. In result, it is impossible to obtain the enough enhancement of $a_{s\ell}^s$ to explain the dimuon charge asymmetry within $1\sigma$ in case i). 

For the cases ii) and iv), the value of $\tilde{h}_s$ takes the dominant role in enhancing the $a_{s\ell}^s$. ($\tilde{h}_s$ must be at least $\mathcal{O}(1)$.) Therefore, the order of couplings $g_{sb}^{L,R} g_{cc}^{L,R}$ must be larger than $10^{-3}$ as our numerical result in Sec. \ref{sec:gcc}. For the case ii), the constraint from $\sin2\beta$ in Sec. \ref{sec:sin2b} excludes the parameter space of $|g_{sb}^{L,R} g_{cc}^{L,R}| > \mathcal{O}(10^{-3})$ just as our result in Sec. \ref{sec:gcc}. In addition, we do not need to consider the case iv) since $|g_{cc}^{L,R}|>1$ to satisfy $|g_{sb}^{L,R} g_{cc}^{L,R}| > \mathcal{O}(10^{-3})$.   

For the case iii), we cannot directly apply the constraint from $\sin2\beta$ since the condition $\theta_L \approx \theta_R$ is not necessary. We now apply $-a_{s\ell}^s/(a_{s\ell}^s)^{\rm SM} > 68$ to Eq. (\ref{eq:asanal}) so that the value of $-\sin\phi_M^s = -$ Arg.($M_{12}^s/M_{12}^{s\,\rm SM}$) must be of order $68  (\Delta M_s / \Delta M_s^{\rm SM}) \sin\phi_s^{\rm SM} \gtrsim 0.2$ from Eq. (\ref{eq:asanal}) since $\Gamma_{12}^{\rm NP}/\Gamma_{12}^{\rm SM}$ is negligible. If we demand a fine cancellation in  (\ref{eq:jppanal}), the value of $\sin\phi_M^s \approx h_s \sin2\sigma_s$ must be similar to $(1.0 \times 10^3)|(g_{cc}^L+g_{cc}^R)(g_{sb}^L-k_\lambda g_{sb}^R)\sin\varphi_\lambda|$, which is $\lesssim |(g_{cc}^L+g_{cc}^R)\sin\varphi_\lambda|$. Therefore, one of the couplings $|g_{cc}^{L,R}|$ must be at least of order $\mathcal{O}(10^{-1})$. Then, we see that the value of $\tilde{h}_s$ is as small as $\mathcal{O}(10^{-2})$ from Eq. (\ref{eq:thsgcc}). Now,  simply assuming $\tilde{h}_s \approx 0$, we can directly compare the experimental bounds of $\Delta M_s$ and $\Delta \Gamma_s$ analytically. As seen in Fig. \ref{fig:tauhst2} (b), the NP contribution in $B_d$ mixing is necessary to explain the dimuon charge asymmetry within $1\sigma$. This result is same as ours in the numerical analysis with the assumption $h_s \approx 0$ in Sec. \ref{sec:gcc}.


\section{Calculation of the contribution to $b \to s\nu\bar{\nu}$}
\label{appen:nu}

All the observables depend on the complex Wilson coefficient $C_L$ and $C_R$ such that
\begin{eqnarray}
{\rm Br}(B \to K^{\ast} \nu \bar{\nu}) &=& 6.8 \times 10^{-6} \left(1+1.31 \eta \right) \epsilon^2~, \\
{\rm Br}(B \to K \nu \bar{\nu}) &=& 4.5 \times 10^{-6} \left(1-2 \eta \right) \epsilon^2~, \\
{\rm Br}(B \to X_s \nu \bar{\nu}) &=& 2.7 \times 10^{-5} \left(1+0.09 \eta \right) \epsilon^2~,
\end{eqnarray}
where $\epsilon$ and $\eta$ are
\begin{eqnarray}
\epsilon &=& \frac{\sqrt{|C_L|^2+|C_R|^2}}{|(C_L)^{\rm SM}|}~, \\
\eta &=& \frac{-{\rm Re} (C_L C_R^{\ast})}{|C_L|^2+|C_R|^2}~.
\end{eqnarray}
The Wilson coefficients are
\begin{eqnarray}
C_L &=& (C_L)^{\rm SM} + (C_L)^{\rm NP} \equiv (C_L)^{\rm SM} - \frac12 \frac{1}{\frac{\alpha}{2\pi} V_{tb} V_{ts}^{\ast}} \frac{1}{g_1^2} \frac{M_Z^2}{M_{Z'}^2} g_{\nu \nu} g_{sb}^L~,  \\
C_R &=& - \frac12 \frac{1}{\frac{\alpha}{2\pi} V_{tb} V_{ts}^{\ast}} \frac{1}{g_1^2} \frac{M_Z^2}{M_{Z'}^2} g_{\nu \nu} g_{sb}^R~,
\end{eqnarray}
where $(C_L)^{\rm SM} = -6.83 \pm 0.06$ 
It is clearly seen that the SM prediction is obtained when $\eta = 0$ and $\epsilon=1$.



\begin{thebibliography}{99}

\def\prp#1#2#3{Phys.\ Rep.\ {\bf #1} (#3) #2}
\def\rmp#1#2#3{Rev. Mod. Phys.\ {\bf #1} (#3) #2}
\def\anrnp#1#2#3{Annu. Rev. Nucl.
Part. Sci.\ {\bf #1} (#3) #2}
\def\npb#1#2#3{Nucl.\ Phys.\ {\bf B\,#1} (#3) #2}
\def\plb#1#2#3{Phys.\ Lett.\ {\bf B\,#1} (#3) #2}
\def\prd#1#2#3{Phys.\ Rev.\ {\bf D\,#1}, #2 (#3)}
\def\prl#1#2#3{Phys.\ Rev.\ Lett.\ {\bf #1} (#3) #2}
\def\jhep#1#2#3{J. High Energy Phys.\ {\bf #1} (#3) #2}
\def\jcap#1#2#3{J. Cos. Astropart. Phys.\ {\bf #1} (#3) #2}
\def\zp#1#2#3{Z.\ Phys.\ {\bf #1} (#3) #2}
\def\epjc#1#2#3{Euro. Phys. J.\ {\bf #1} (#3) #2}
\def\ijmp#1#2#3{Int.\ J.\ Mod.\ Phys.\ {\bf #1} (#3) #2}
\def\mpl#1#2#3{Mod.\ Phys.\ Lett.\ {\bf #1} (#3) #2}
\def\apj#1#2#3{Astrophys.\ J.\ {\bf #1} (#3) #2}
\def\nat#1#2#3{Nature\ {\bf #1} (#3) #2}
\def\sjnp#1#2#3{Sov.\ J.\ Nucl.\ Phys.\ {\bf #1} (#3) #2}
\def\apj#1#2#3{Astrophys.\ J.\ {\bf #1} (#3) #2}
\def\ijmp#1#2#3{Int.\ J.\ Mod.\ Phys.\ {\bf #1} (#3) #2}
\def\apph#1#2#3{Astropart.\ Phys.\ {\bf B\,#1}, #2 (#3)}


\bibitem{Abazov:2010hv}
  V.~M.~Abazov {\it et al.} [ D0 Collaboration ],
  Phys.\ Rev.\  {\bf D82}, 032001 (2010)
  [arXiv:1005.2757 [hep-ex]].

\bibitem{Abazov:2011yk} 
  V.~M.~Abazov {\it et al.}  [D0 Collaboration],
  Phys.\ Rev.\ D {\bf 84}, 052007 (2011)
  [arXiv:1106.6308 [hep-ex]].

\bibitem{Lenz:2011ti}
  A.~Lenz and U.~Nierste,
  [arXiv:1102.4274 [hep-ph]].
   

\bibitem{Kim:2010gx}
  J.~E.~Kim, M.~-S.~Seo and S.~Shin,
  Phys.\ Rev.\  {\bf D83}, 036003 (2011)
  [arXiv:1010.5123 [hep-ph]].
  
\bibitem{Alok:2010ij}
  A.~K.~Alok, S.~Baek and D.~London,
  JHEP {\bf 1107}, 111 (2011)
  [arXiv:1010.1333 [hep-ph]].  



\bibitem{Langacker:2000ju} 
  P.~Langacker and M.~Plumacher,
  Phys.\ Rev.\ D {\bf 62}, 013006 (2000)
  [hep-ph/0001204].
  

\bibitem{Kim:2011xv} 
  J.~E.~Kim and S.~Shin,
  Phys.\ Rev.\ D {\bf 85}, 015012 (2012)
  [arXiv:1104.5500 [hep-ph]].


\bibitem{Bauer:2010dga}
  C.~W.~Bauer and N.~D.~Dunn,
  Phys.\ Lett.\  {\bf B696}, 362-366 (2011) 
  [arXiv:1006.1629 [hep-ph]].
  
\bibitem{Li:2012xc} 
  X.~-Q.~Li, Y.~-M.~Li, G.~-R.~Lu and F.~Su,
  JHEP {\bf 1205}, 049 (2012)
  [arXiv:1204.5250 [hep-ph]].
 

  
\bibitem{Lenz:2006hd}
  A.~Lenz and U.~Nierste,
  JHEP {\bf 0706}, 072 (2007) 
  [hep-ph/0612167]. 

\bibitem{LHCb-CONF-2011-050}
 LHCb-CONF-2011-050

\bibitem{Barberio:2008fa}
  E.~Barberio {\it et al.} [ Heavy Flavor Averaging Group Collaboration ],
  [arXiv:0808.1297 [hep-ex]].

  
\bibitem{Dermisek:2011xu}
  R.~Dermisek, S.~-G.~Kim and A.~Raval,
  Phys.\ Rev.\ D {\bf 84}, 035006 (2011)
  [arXiv:1105.0773 [hep-ph]] ;
  R.~Dermisek, S.~-G.~Kim and A.~Raval,
  Phys.\ Rev.\ D {\bf 85}, 075022 (2012)
  [arXiv:1201.0315 [hep-ph]].
  
  
\bibitem{Acosta:2005ij} 
  D.~Acosta {\it et al.}  [CDF Collaboration],
  Phys.\ Rev.\ Lett.\  {\bf 95}, 131801 (2005)
  [hep-ex/0506034].
  
\bibitem{Chatrchyan:2012hd} 
  S.~Chatrchyan {\it et al.}  [CMS Collaboration],
  Phys.\ Lett.\ B {\bf 716}, 82 (2012)
  [arXiv:1206.1725 [hep-ex]]. 
    


\bibitem{Inami:1980fz} 
  T.~Inami and C.~S.~Lim,
  Prog.\ Theor.\ Phys.\  {\bf 65}, 297 (1981)
  [Erratum-ibid.\  {\bf 65}, 1772 (1981)].

\bibitem{Buchalla:1995vs} 
  G.~Buchalla, A.~J.~Buras and M.~E.~Lautenbacher,
  Rev.\ Mod.\ Phys.\  {\bf 68}, 1125 (1996)
  [hep-ph/9512380].


\bibitem{PDG2011}
   K. Nakamura et al. (Particle Data Group), Journal of Physics G37, 075021 (2010) and 2011 partial update for the 2012 edition.

\bibitem{Lancaster:2011wr} 
  [Tevatron Electroweak Working Group and CDF and D0 Collaborations],
  arXiv:1107.5255 [hep-ex].
  
\bibitem{Lenz:2012az} 
  A.~Lenz, U.~Nierste, J.~Charles, S.~Descotes-Genon, H.~Lacker, S.~Monteil, V.~Niess and S.~T'Jampens,
  Phys.\ Rev.\ D {\bf 86}, 033008 (2012)
  [arXiv:1203.0238 [hep-ph]].  
  
\bibitem{Buras:2001ra} 
  A.~J.~Buras, S.~Jager and J.~Urban,
  Nucl.\ Phys.\ B {\bf 605}, 600 (2001)
  [hep-ph/0102316].  
    
  
\bibitem{Becirevic:2001xt} 
  D.~Becirevic, V.~Gimenez, G.~Martinelli, M.~Papinutto and J.~Reyes,
  JHEP {\bf 0204}, 025 (2002)
  [hep-lat/0110091].
  

\bibitem{lhcbn1}
 LHCb-CONF-2012-002
  
\bibitem{Chiang:2009ev}
  C.~-W.~Chiang, A.~Datta, M.~Duraisamy, D.~London, M.~Nagashima and A.~Szynkman,
  JHEP {\bf 1004}, 031 (2010)
  [arXiv:0910.2929 [hep-ph]].  
  
\bibitem{d0phi}
 S. Burdin [D0 Collaboration], ``Measurements of CP violation in the $B_s$ system at D0", talk given at the Europhysics Conference on High-Energy Physics 2011, July 21, 2011. 
  
\bibitem{Melikhov:2000yu} 
  D.~Melikhov and B.~Stech,
  Phys.\ Rev.\ D {\bf 62}, 014006 (2000) 
  [hep-ph/0001113]. 
  
\bibitem{Ball:2004rg} 
  P.~Ball and R.~Zwicky,
  Phys.\ Rev.\ D {\bf 71}, 014029 (2005)
  [hep-ph/0412079].


 \bibitem{Bobeth:2011st}
  C.~Bobeth and U.~Haisch,
  [arXiv:1109.1826 [hep-ph]]. 
  
\bibitem{Altmannshofer:2009ma}
  W.~Altmannshofer, A.~J.~Buras, D.~M.~Straub, M.~Wick,
  JHEP {\bf 0904}, 022 (2009) 
  [arXiv:0902.0160 [hep-ph]]. 
   
   
\bibitem{arXiv:1111.1257} 
  W.~Altmannshofer, P.~Paradisi and D.~M.~Straub,
  JHEP {\bf 1204}, 008 (2012)
  [arXiv:1111.1257 [hep-ph]].
  
\bibitem{hep-ex/0010022} 
  R.~Barate {\it et al.} [ALEPH Collaboration],
  Eur.\ Phys.\ J.\ C\ {\bf 19}, 213  (2001) 
  [hep-ex/0010022].     

\bibitem{Fox:2011qd} 
  P.~J.~Fox, J.~Liu, D.~Tucker-Smith and N.~Weiner,
  Phys.\ Rev.\ D {\bf 84}, 115006 (2011)
  [arXiv:1104.4127 [hep-ph]].

\bibitem{Charles:2004jd}
  J.~Charles {\it et al.} [ CKMfitter Group Collaboration ],
  Eur.\ Phys.\ J.\  {\bf C41}, 1-131 (2005)
  [hep-ph/0406184].   

\bibitem{demiseCKM}
  E.~Lunghi and A.~Soni,
  arXiv:1104.2117 [hep-ph];
  E.~Lunghi and A.~Soni,
  Phys.\ Lett.\  B {\bf 697}, 323 (2011)
  [arXiv:1010.6069 [hep-ph]].

\bibitem{Adachi:2012mm} 
  I.~Adachi {\it et al.}  [Belle Collaboration],
  arXiv:1208.4678 [hep-ex].

\bibitem{Golowich:2006gq}
  E.~Golowich, S.~Pakvasa and A.~A.~Petrov,
  Phys.\ Rev.\ Lett.\  {\bf 98}, 181801 (2007)
  [hep-ph/0610039].

\bibitem{Chen:2007dg}
  S.~-L.~Chen, X.~-G.~He, A.~Hovhannisyan {\it et al.},
  JHEP {\bf 0709}, 044 (2007)
  [arXiv:0706.1100 [hep-ph]].

\bibitem{Zupanc:2007pu} 
  A.~Zupanc, K.~Abe, K.~Abe, H.~Aihara, D.~Anipko, K.~Arinstein, V.~Aulchenko and T.~Aushev {\it et al.},
  Phys.\ Rev.\ D {\bf 75}, 091102 (2007)
  [hep-ex/0703040].

\bibitem{Giudice:2012qq} 
  G.~F.~Giudice, G.~Isidori and P.~Paradisi,
  JHEP {\bf 1204}, 060 (2012)
  [arXiv:1201.6204 [hep-ph]].
  
\bibitem{Altmannshofer:2012ur} 
  W.~Altmannshofer, R.~Primulando, C.~-T.~Yu and F.~Yu,
  JHEP {\bf 1204}, 049 (2012)
  [arXiv:1202.2866 [hep-ph]].
    

\bibitem{Acosta:2004gt}
  D.~Acosta {\it et al.}  [CDF Collaboration],
  Phys.\ Rev.\ Lett.\  {\bf 94}, 101803 (2005)
  [arXiv:hep-ex/0412057].  

\bibitem{hep-ph/0609232} 
  M.~Misiak, H.~M.~Asatrian, K.~Bieri, M.~Czakon, A.~Czarnecki, T.~Ewerth, A.~Ferroglia and P.~Gambino {\it et al.},
  Phys.\ Rev.\ Lett.\ \ {\bf 98}, 022002  (2007)
  [hep-ph/0609232].
  
\bibitem{arXiv:1010.1589} 
  D.~Asner {\it et al.} [Heavy Flavor Averaging Group Collaboration],
  arXiv:1010.1589 [hep-ex].  


\bibitem{Buras:2011zb}
  A.~J.~Buras, L.~Merlo and E.~Stamou,
  JHEP {\bf 1108} (2011) 124
  [arXiv:1105.5146 [hep-ph]].






\end{thebibliography}
\end{document}